\documentclass[prb,aps,amsmath,amssymb,showpacs,twocolumn,notitlepage]{revtex4-1}
\usepackage{amssymb,graphicx,float,dcolumn,bm,subfigure,color}
\usepackage[normalem]{ulem}
\usepackage[titletoc,title]{appendix}
\usepackage[dotinlabels]{titletoc}
\usepackage{subfigure}
\usepackage{braket}
\usepackage{nicefrac}
\usepackage[dvipsnames]{xcolor}

\newcommand{\be}{\begin{equation}}
\newcommand{\ee}{\end{equation}}

\newcommand{\ba}{\begin{eqnarray}}
\newcommand{\ea}{\end{eqnarray}}

\renewcommand{\vec}[1]{{\textbf{\textit{#1}}}}

\begin{document}
\title{Kohn-Sham Density Functional Theory of Abelian Anyons}
\author{Yayun Hu,$^1$ G. Murthy,$^2$ S. Rao,$^3$ J. K. Jain$^1$}
\affiliation{$^1$ Physics Department, 104 Davey Lab, Pennsylvania State University, University Park, Pennsylvania, 16801, USA}
\affiliation{$^2$Department of Physics and Astronomy, University of Kentucky, Lexington, KY 40506-0055}
\affiliation{$^3$Harish-Chandra Research Institute, HBNI, Chhatnag Road, Jhunsi, Allahabad 211 019, India}

\begin{abstract}
We develop a density functional treatment of non-interacting abelian anyons, which is capable, in principle, of dealing with a system of a large number of anyons in an external potential. Comparison with exact results for few particles shows that the model captures the behavior qualitatively and semi-quantitatively, especially in the vicinity of the fermionic statistics. We then study anyons with statistics parameter $1+1/n$, which are thought to condense into a superconducting state. An indication of the superconducting behavior is the mean-field result that, for uniform density systems, the ground state energy increases under the application of an external magnetic field independent of its direction. Our density-functional-theory based analysis does not find that to be the case for finite systems of anyons, which can accommodate a weak external magnetic field through density transfer between the bulk and the boundary rather than through transitions across effective Landau levels, but the ``Meissner repulsion" of the external magnetic field is recovered in the thermodynamic limit as the effect of the boundary becomes negligible.  We also consider the quantum Hall effect of anyons, and show that its topological properties, such as the charge and statistics of the excitations and the quantized Hall conductance, arise in a self-consistent fashion. 
\end{abstract}

\maketitle

\tableofcontents

\section{Introduction}

Anyons are particles obeying fractional braiding statistics~\cite{Leinaas77,Wilczek82}. Anyons with statistics parameter $1+\alpha$ are conveniently modeled as fermions carrying flux of magnitude $\alpha\phi_0$, where $\phi_0=h/e$ is the flux quantum~\cite{Wilczek90,Roberto92,Lerda92,Rao92,Khare05,Rao17}.
An exchange of these anyons produces a phase factor $(-1)^{1+\alpha}$, interpolating between bosons ($\alpha=-1$) and fermions ($\alpha=0$) in two dimensions. Anyons have been treated by field theoretical methods\cite{Chen89,Roberto91,Roberto92}, exact diagonalization\cite{Sporre91,Murthy91,Sporre92,Sporre93,Canright89,Canright89a,Hanna89,Xie90,Hatsugai91,Kudo20}, construction of wave functions~\cite{Wu84,Girvin90,Chin92,Lundholm17,Lundholm13}, and other approaches~\cite{Laughlin88science,Fetter89,Wen90b,Lee91,Chitra92,Li92,Correggi17}. The aim of this article is to develop a Kohn-Sham density-functional theory (KS-DFT) for anyons, which implements the binding of flux quanta to electrons through a density dependent effective magnetic field within a self-consistent KS scheme. The motivation for developing the KS-DFT for anyons is two-fold: (i) it allows us to deal with anyon ground states with inhomogeneous densities, as is the case for anyons in an external potential; and (ii) it enables a study of systems with a large number of anyons. Because we are most interested in topological features, we consider in this work the Kohn-Sham theory without any exchange-correlation energy. (The flux bound to the anyons effectively generates a long range gauge interaction, which will produce an exchange-correlation energy even for non-interacting anyons. This energy is relevant for the non-topological features of the solution. However, we will be most interested in the vicinity of the fermionic statistics, i.e. small $\alpha$, where this contribution is expected to be weak.) 

The study of anyons in confined systems is particularly important and timely because of recent experimental reports of fractional statistics in the fractional quantum Hall effect (FQHE)~\cite{Tsui82} in interference and collision experiments~\cite{Nakamura20,Bartolomei20}, consistent with theoretical expectations~\cite{Halperin84,Arovas85}. The FQHE at fractions of the form $\nu=n/(2pn\pm 1)$ is an integer quantum Hall effect of emergent particles called composite fermions (CFs)~\cite{Jain89,Jain07,Halperin20}. Composite fermions are closely related to anyons, in that they are also often modeled as electrons carrying gauge flux quanta, although in this case, the number of flux quanta bound to electrons is an even integer $2p$, which brings the statistics back to fermionic. [The quasiparticles of the FQHE state are excited composite fermions. These, however, carry a fractional charge, which leads to fractional braiding statistics, as verified in explicit calculations~\cite{Kjonsberg99b,Jeon03b,Jeon04,Jain89,Jain07}.]

The present work closely follows a recently developed KS-DFT to deal with a system of composite fermions~\cite{Hu19}. 
We stress that we are dealing here with a system of anyons without reference to their origin; for the quasiparticles of the FQHE it would actually be more appropriate to use the KS theory of composite fermions~\cite{Hu19}, which also includes an exchange-correlation energy and thereby captures quantitative features beyond their charge and fractional braiding.

We summarize here the principal findings of the present work. 

We first consider systems of three and four anyons for which exact solutions are available~\cite{Sporre91,Murthy91,Sporre92,Sporre93}. We find that the KS-DFT method captures the behavior of energy as a function of $\alpha$ in the vicinity of the fermion statistics (i.e., for small $\alpha$). We also compare DFT against a trial wave function of the kind used for composite fermions, and find that the results from the two approaches are similar near the fermionic point. Other trial wave functions for anyons can be found in the literature~\cite{Girvin90,Chin92,Lundholm17,Lundholm13,Ouvry07,Lundholm13a,Ilieva01},  but are not considered here.

We next investigate how the nature of the confinement potential affects the behavior. To that end, we study  two external potentials, the parabolic potential and the infinite hard wall potential. These two potentials shows distinct behaviors in how the total energy and density distribution of anyons scale with particle number $N$ in a large system. We note that a semiclassical Thomas-Fermi approximation has also been used to treat anyons~\cite{Chitra92,Li92,Correggi17}. Some of the features from this method, such as the density and the energy in the presence of an external potential, are qualitatively and semi-quantitatively similar to those  found from our DFT method. The DFT method, of course, suggests a new scheme for improving the results, and most importantly, allows a treatment of the topological properties.

We then address the proposal that anyons with $\alpha=1/n$ exhibit superconductivity~\cite{Laughlin88science,Fetter89,Chen89,Canright89,Canright89a,Lee89,Hanna89,Xie90,Mcmullen91,Lee91,Kudo20}. At a mean field level, where the density is assumed to be uniform, such anyons correspond to fermions filling $n$ effective Landau levels (LLs), the energy of which increases upon application of a magnetic field, independent of its direction. There are actually two competing effects: in one direction, the total ``effective" magnetic field of the mean field theory increases, which leads to a higher cyclotron energy but at the same time some particles move to lower LLs; in the other direction, the cyclotron energy is reduced but some particles are pushed to higher LLs~\cite{Chen89}. Combining the two effects reveals that the net energy increases for either direction of the externally applied magnetic field, which is taken as a signature of the Meissner effect~\cite{Chen89,Pierre99}. 

Our KS-DFT calculations find that, for a finite system of anyons in a confining potential, the application of an external magnetic field increases the energy in one direction but not the other. 
We explain this as arising from the fact that, because of the presence of boundaries, the systems can accommodate a small external magnetic field by an adjustment of the density profile without requiring transitions between the effective LLs. 
Nonetheless, we find that the expected ``Meissner expulsion" of the external magnetic field is recovered in the thermodynamic limit, as the importance of edge effects diminishes.  More specifically, we find that the energy has a minimum at a small external magnetic field, denoted $B_{\rm min}$, and that $B_{\rm min}$ vanishes in the thermodynamic limit. 

We also study the response due to a localized magnetic flux for $\alpha=1/n$, and find that it scales with the number of particles. We take that to be consistent with anyon superconductivity.

Finally, we consider the integer quantum Hall effect (IQHE) of anyons. Many properties of this state, such as Hall resistance and the charge and the statistics of the quasiparticles, can be derived in a manner analogous to the IQHE of composite fermions. We find that the charge and statistics of the emergent quasi-particles are different from those of the original anyons forming the IQHE state. This is unlike what happens for fermions  in the IQHE, but similar to what happens for composite fermions in the FQHE state. This is not surprising, as even free anyons form a correlated state. The charge and statistics of the excitations for certain IQHE states, which are amenable to the plasma analogy, were calculated previously~\cite{Ma91}.

The plan of our paper is as follows. The DFT formulation of anyons is introduced in Sec.~\ref{KSsection}. We compare the DFT results with those obtained using exact diagonalization and trial wave functions  in Sec.~\ref{Numerical190719} and Sec.~\ref{Trialsection}. We then proceed to study a large number anyons in a confining potential in Sec.~\ref{Largesection}, where we consider parabolic as well as hard wall confinement. Finally, we consider the IQHE of anyons using the DFT method in Sec.~\ref{topogicalQH}. The paper is concluded with some discussions in Sec.~\ref{sec:Conclusions}.

\section{Kohn-Sham equations for non-interacting anyons}\label{KSsection}

Let us begin with a brief background of the KS-DFT approach for FQHE. The KS formulation of the DFT~\cite{Hohenberg64,Kohn65} is a very useful method for dealing with inhomogeneous systems of interacting particles (see Ref.~\onlinecite{Giuliani08}). This theory uses the electron density to reduce the many particle problem to a single particle problem and, in principle, gives the exact ground state density and energy provided the so-called universal ``exchange correlation function" is known. All of the complexities of the interaction are hidden in the exchange correlation function. The KS-DFT works best when the kinetic energy provides the dominant contribution. One of the problems in applying the DFT to the FQHE~\cite{Ferconi95,Heinonen95} is that the FQHE state is a non-perturbative state whose physics is entirely determined by the interaction energy. From another perspective, the FQHE ground state cannot be approximated by a single Slater determinant ground state of non-interacting particles in a KS potential; in the latter, the KS orbitals are either occupied or unoccupied, whereas in the FQHE state the KS orbitals are fractionally occupied. Ref.~\onlinecite{Zhao17} proposed to define the DFT in terms of composite fermions, which addressed some of the difficulties. 
Ref.~\onlinecite{Hu19} further developed the KS-DFT for composite fermions, which properly accounts for the flux quanta bound to composite fermions and captures the topological properties of the FQHE, such as the fractional charge~\cite{Laughlin83} and fractional statistics~\cite{Halperin84} of its excitations. The effect of the Berry phases produced by the bound flux quanta is treated in this approach through a non-local interaction induced by the gauge field. The KS-DFT approach is also able to deal with systems with inhomogeneous densities. Here we apply an  analogous KS-DFT to a system of noninteracting anyons.

\subsection{Anyons as electrons with attached flux tubes}\label{HKandFQHE}

We consider a system in a 2D plane perpendicular to the unit vector $\vec{e}_{\rm z}$ along the $z$ axis. The Hamiltonian 
for anyons is:
\be
\mathcal H_{A}\Psi_{A}=E\Psi_{A},
\label{AnyonSchrodingerEq}
\ee
with the exchange boundary condition 
\begin{eqnarray}
\Psi_{A}(\vec{r}_1,\ldots,\vec{r}_i,\ldots,\vec{r}_j,\ldots,\vec{r}_N)=\nonumber
\\ (-1)^{1+\alpha}\Psi_{A}(\vec{r}_1,\ldots,\vec{r}_j,\ldots,\vec{r}_i,\ldots,\vec{r}_N),
\label{AnyonBoundaryCondition}
\end{eqnarray}
where $\vec{r}_i=(x_i,y_i)$ is the position of the $i^{\rm th}$ anyon and $\alpha$ is the statistical parameter. The statistics becomes fermionic (bosonic) when $\alpha \mod 2=0 (1)$. For anyons in an external vector potential $\vec{A}_{\rm ext}$ as well as a scalar potential $V_{\rm ext}$, the Hamiltonian is
\be
\mathcal H_{A}=\sum_i\left[\frac{1}{2M}\left(\vec{p}_i-\frac{q}{c}\vec{A}_{\rm ext}(\vec{r}_i)\right)^2+V_{\rm ext}(\vec{r}_i)\right],
\label{AnyonHamiltonian}
\ee
where $q$ is the charge of each anyon, and we have assumed that there is no interaction between anyons.
We note that unlike for bosons or fermions, even the problem of noninteracting anyons is nontrivial, and does not lend itself to an exact solution.

Next we follow the standard procedure and make a singular Chern-Simons transformation 
\be
\Psi_{A}=\prod_{j<k}\left(\frac{z_j-z_k}{|z_j-z_k|}\right)^{\alpha}\Psi_{F}=\exp(i\alpha\sum_{j<k}\theta_{jk})\Psi_{F},
\ee 
where $\theta_{jk}=-i\ln{\frac{z_j-z_k}{|z_j-z_k|}}$, and $z_j=x_j+iy_j$ is the complex position. The transformed problem for fermions is given by
\be
\mathcal H_F \Psi_F = E \Psi_F
\label{FermionSchrodingerEq}
\ee
where $\Psi_F$ is antisymmetric under coordinate exchange, and  
\be
\mathcal H_F=\sum_i\left[\frac{1}{2M}\left(\vec{p}_i-\frac{q}{c}\vec{A}_{\rm ext}(\vec{r}_i)+\frac{e}{c}\vec{a}_i\right)^2+V_{\rm ext}(\vec{r}_i)\right],
\label{FullHF}
\ee
with Chern-Simons vector potential $\vec{a}_i$ given by
\be
\vec{a}_i=\frac{\alpha\phi_0}{2\pi}\sum_{j(j\neq i)}{\vphantom{\sum}}\nabla_i\theta_{ij}=\frac{\alpha\phi_0}{2\pi}\sum_{j(j\neq i)}{\vphantom{\sum}}\left(\frac{y_j-y_i}{r^2_{ij}},\frac{x_i-x_j}{r^2_{ij}}\right).
\label{vortexpotential}
\ee
The boundary condition in Eq.\ref{AnyonSchrodingerEq} is now replaced by the statistical interaction between particles  via the $\vec{a}$ term, because of which they see $\alpha$ vortices on one another:
\be
\nabla_i\times\vec{a}_i(\vec{r}_i)=\alpha\phi_0\sum_{j(j\neq i)}{\vphantom{\sum}}\delta^{(2)}(\vec{r}_i-\vec{r}_j)\vec{e}_{\rm z}.
\label{vortexfield}
\ee
The problem of non-interacting anyons thus maps into a problem of fermions with very complex long range interaction induced by the braid statistics. 
We note that we also have the choice to map the anyonic problem into a bosonic one~\cite{Canright89a,Lee89,Diptiman92b,Correggi17}, but the fermionic representation is more convenient because it automatically eliminates coincidences of particles~\cite{Hanna89,Mancarella13}.

According to the generalized Hohenberg-Kohn (HK) theorem in an external magnetic field, i.e., the so-called magnetic-field DFT~\cite{Grayce94,kohn04,Tellgren12,Tellgren18b}, the ground state energy of a system of fermions in a magnetic field is a functional of the density. We will assume that to be the case for the system in Eq.~\ref{FermionSchrodingerEq} (see Appendix~\ref{GeneralHK} for details) and write
\be
E^{\alpha}[\rho, \vec{A}_{\rm ext}]=F^{\alpha}[\rho, \vec{A}_{\rm ext}]+\int d\vec{r} V_{\rm ext}(\vec{r})\rho(\vec{r}),
\label{DFT12}
\ee
where the functional $F^{\alpha}$ is defined through a constrained search~\cite{Levy79,Lieb83}
\be\label{KSstartotalEK}
F^{\alpha}[\rho, \vec{A}_{\rm ext}]=\min_{\Psi\rightarrow \rho}\langle\Psi | \sum_i\frac{\left(\vec{p}_i-\frac{q}{c}\vec{A}_{\rm ext}(\vec{r}_i)+\frac{e}{c}\vec{a}_i\right)^2}{2M}| \Psi\rangle,
\ee
where the minimum is obtained by searching over all fermionic wave functions that reproduce the density $\rho$. Notice that the energy $F^{\alpha}$ is a functional of both the density and the external vector potential. This should be contrasted with another version of DFT in an external magnetic field, dubbed the current-density functional theory (CDFT)~\cite{Vignale87}; in this approach, the universal energy functional depends explicitly on density, current distribution and magnetization, but has no dependence on the vector potential. 

For reasons of convergence, we will often work with a small finite temperature. At temperature $\tau$, $F^{\alpha}$ is generalized to a constrained search over all density matrices $\hat{\Gamma}$ that reproduce $\rho$, 
\begin{eqnarray}
&&F^{\alpha}[\rho, \vec{A}_{\rm ext}, \tau]=\nonumber\\ 
&&\min_{\hat{\Gamma}\rightarrow \rho}\textrm{Tr}~\hat{\Gamma}\left[\sum_i\frac{\left(\vec{p}_i-\frac{q}{c}\vec{A}_{\rm ext}(\vec{r}_i)+\frac{e}{c}\vec{a}_i\right)^2}{2M}-k_{\rm B}\tau\hat{S}\right]\;\label{RealFunctionalT},
\end{eqnarray}
where $\hat{S}$ is the entropy operator\cite{Jones14}.

\subsection{Auxiliary KS system}\label{KSstarSection}

We next proceed to define an auxiliary KS system of ``non-interacting" fermions, in the same way as previously done for composite fermions in the fractional quantum Hall effect\cite{Hu19}. The KS Hamiltonian is defined as
\begin{equation}\label{HStar}
\mathcal H_{\rm KS}[\rho]=\sum_j\frac{1}{2M}\left(\vec{p}_j-\frac{q}{c}\vec{A}_{\rm ext}(\vec{r})+\frac{e}{c}\vec{A}_{\rm eff}(\vec{r}_j)\right)^2+{V}_{\rm KS}(\vec{r}),
\end{equation}
with
\begin{eqnarray}
  \nabla \times \vec{A}_{\rm ext}(\vec{r})=B_{\rm ext}(\vec{r})\vec{e}_{\rm z}\;, \\
  \nabla \times \vec{A}_{\rm eff}(\vec{r})=\alpha\rho(\vec{r})\phi_0\vec{e}_{\rm z}\; ,\label{AequationMain}
\end{eqnarray}
where we consider only external magnetic field perpendicular to the system, and $\rho(\vec{r})$ is the anyon density. From now on, without any loss of generality, we choose the anyon charge $q=-e$; it is equivalent to a different choice for the charge of anyons through a rescaling of the external magnetic field $B_{\rm ext}$~\cite{Choi99}. This way, we have total vector potential $\vec{A}(\vec{r})$: 
\be
\vec{A}(\vec{r})=\vec{A}_{\rm ext}(\vec{r})+\vec{A}_{\rm eff}(\vec{r})\;,\label{Acomposition}
\ee
and total magnetic field $\vec{B}(\vec{r})$:
\begin{equation}
\vec{B}(\vec{r})=\left[B_{\rm ext}(\vec{r})+\alpha\rho(\vec{r})\phi_0\right]\vec{e}_{\rm z}\;,\label{AequationMain}
\end{equation}
which is a sum of the external magnetic field $B_{\rm ext}\vec{e}_{\rm z}$ and the effective magnetic field $\vec{B}_{\rm eff}=\alpha\rho(\vec{r})\phi_0\vec{e}_{\rm z}$ generated by the flux attached to the anyons.

For rotationally symmetric systems, which are what we consider in this article, it is convenient to choose the gauge
\be
\vec{A}(\vec{r})=\frac{r\mathcal{B}(r)}{2}\vec{e}_{\rm \phi}\;,
\label{Ar*}
\ee
with
\begin{equation}
  \mathcal{B}(r)=\frac{1}{\pi r^2}\int_0^r 2\pi r'B(r') dr'\;.\label{pseudoB}
\end{equation}
The KS orbitals $\psi_a(\vec{r})$, where $a$ collectively denotes the quantum numbers, are the solutions of  
\be
\left[ T +V_{\rm KS}(\vec{r})  \right] \psi_{a}(\vec{r}) = \epsilon_{a} \psi_{a}(\vec{r})\;
\label{singleCFKS}
\ee
with
\begin{equation}
  T=\frac{1}{2M}\left(\vec{p}+\frac{e}{c}\vec{A}(\vec{r})\right)^2. \label{SingleParticleTstar}
\end{equation}
The fermion density is given by
\be
\rho(\vec{r})=\sum_{a} c_{a}|\psi_{a}(\vec{r})|^2\;,
\label{rho*rho}
\ee
where $c_{a}=1/[e^{(\epsilon_{a}-\mu)/k_{\textrm{B}}\tau}+1]$ is the occupation number of the KS orbital labeled by $a$, and the chemical potential $\mu$ is determined by the condition $\sum_ac_a=N$.

We stress that even the problem of seemingly ``noninteracting" particles in Eq.~\ref{HStar} is not really noninteracting, as evident from the fact that the vector potential depends on the density. The solution for the KS problem must be obtained self-consistently, and the wave function and the energy of each KS orbital depend on which other KS orbitals are occupied.

In the auxiliary KS system, we define the so-called kentropy $K^{\alpha}_{\rm s}$ as the sum of the non-interacting kinetic energy and the non-interacting entropy:
\begin{eqnarray}
K^{\alpha}_{\rm s}[\rho, \vec{A}_{\rm ext}, \tau]&=\sum_{a=1}^Nc_a\langle\psi_a | \frac{1}{2M}\left(\vec{p}+\frac{e}{c}\vec{A}(\vec{r})\right)^2 | \psi_a\rangle\nonumber\\&-k_{\textrm{B}}\tau\sum_a c_a\log(c_a)\;.
\end{eqnarray}
At zero temperature, where $c_a=1$ for occupied states and $c_a=0$ for unoccupied states, $K^{\alpha}_{\rm s}$ reduces to the more familiar $T_s$, which is the kinetic energy of noninteracting particles in the KS potential. The role of $K^{\alpha}_{\rm s}$ becomes evident in the following. 

\subsection{KS equations}\label{KSequationsMap}

Next, following the standard DFT formulation, we map the interacting fermions in Sec.~\ref{HKandFQHE} to a system of auxiliary particles in Sec. \ref{KSstarSection} by making a key assumption:  {\it The density profile of any physically relevant anyon ground state can be obtained in the auxiliary Kohn-Sham problem by an appropriate choice of $V_{\rm KS}(\vec{r})$.} 

To this end, we write the HK functional $F_{\alpha}[\rho]$ in Eq.~\ref{RealFunctionalT} as
\begin{eqnarray}
\label{NewPartitionFrho}
F^{\alpha}[\rho, \vec{A}_{\rm ext}, \tau]=K^{\alpha}_{\rm s}[\rho, \vec{A}_{\rm ext}, \tau]+E^{\alpha}_{\rm xc}[\rho, \vec{A}_{\rm ext}, \tau]\;.
\end{eqnarray}
Such a partitioning of $F^{\alpha}$ can, in principle, always be made for an appropriate choice of the exchange-correlation (xc) energy $E^{\alpha}_{\rm xc}$. We now need to minimize the energy
\begin{eqnarray}
E^{\alpha}[\rho, \vec{A}_{\rm ext}, \tau]&=K^{\alpha}_{\rm s}[\rho, \vec{A}_{\rm ext}, \tau]+E^{\alpha}_{\rm xc}[\rho, \vec{A}_{\rm ext}, \tau]\nonumber\\
&+\int d\vec{r} V_{\rm ext}(\vec{r})\rho(\vec{r})\;,\label{Interpretation}
\end{eqnarray}
for a system with a fixed number of particles with the constraint $\int d \vec{r} \psi_{a}^*(\vec{r})\psi_{b}(\vec{r})=\delta_{ab}$. 
This leads to the KS  Eq.~(\ref{singleCFKS}) with
\begin{equation}\label{VKSstarDisplay}
  V_{\rm KS}[\rho, \vec{A}_{\rm ext}, \tau](\vec{r})=V_{\rm T}(\vec{r})+V^{\alpha}_{\rm xc}(\vec{r})+V_{\rm ext}(\vec{r})\;.
\end{equation}
Here the non-standard potential $V_{\rm T}$: 
\be
V_{\rm T}[\rho, \vec{A}_{\rm ext}, \tau](\vec{r})=\sum_{a}\langle \psi_a|\frac{\delta T}{\delta \rho_{\vec{r}}}|\psi_a\rangle,
\ee
comes from the density dependence of $T$,
and $V^{\alpha}_{\rm xc}[\rho, \vec{A}_{\rm ext}, \tau](\vec{r})=\frac{\delta E^{\alpha}_{\rm xc}}{\delta \rho(\vec{r})}$ is the exchange-correlation potential.

The exchange-correlation potential $V^{\alpha}_{\rm xc}(\vec{r})$, is a functional of density and does not explicitly depend on $V_{\rm ext}(\vec{r})$. Throughout this paper, we make a simplifying assumption of choosing $V_{\rm T}(\vec{r})=V^{\alpha}_{\rm xc}(\vec{r})=0$. This is justified as follows. The interaction between fermions arises solely due to the statistical gauge flux quanta attached to them. The effect of this interaction is included, to some extent, through the effective magnetic field $B_{\rm eff}$. As we see in the problem of $\rm FQHE$~\cite{Hu19},  the effective magnetic field takes care of the long-range gauge interaction, which is non-local and is responsible for several topological properties of the FQHE, which are largely independent of $V_{\rm T}$ as well as the choice of the xc potential in the KS theory of FQHE.  We follow an analogous approach here. The validity of our assumption of dropping $V_{\rm T}(\vec{r})$ and $V^{\alpha}_{\rm xc}(\vec{r})$ will be judged a posteriori by comparison with exact results, but we expect that it will capture some essential features of the anyon system. The KS equation still needs to be solved self-consistently because $T$ depends on the density.

Despite the above simplifying assumptions, there is still rich physics tunable by the particle number and the external magnetic field, as we see below. In the absence of an external magnetic field, Eq.~\ref{pseudoB} yields the vector potential 
\be
\vec{A}(r)=\vec{A}_{\rm eff}(r)=\frac{\alpha\phi_0N(r)}{2\pi r},\; N(r)=\int_0^{r}\rho(r')2\pi r'dr',
\ee
where $N(r)$ is the number of particles inside the region of radius $r$. The kinetic operator in Eq.~\ref{SingleParticleTstar} can now be writen as
\be
T=\frac{1}{2M}\left[\vec{p}+\hbar\alpha\frac{N(r)}{r}\vec{e}_{\phi}\right]^2. \label{TnoBext}
\ee

Among the two confinement potentials considered here, one is the parabolic potential 
\be
V_{\rm para}(\vec{r})=M\omega^2r^2/2. 
\ee
$\mathcal H_{\rm KS}$ can be written in a dimensionless form as
\be
\mathcal H_{\rm KS}=\hbar\omega\left\{  \left[-i\frac{1}{\bar{r}}\frac{\partial}{\partial\phi}+\frac{\alpha N(\bar{r})}{\bar{r}}\right]^2+(-i\frac{\partial}{\partial\bar{r}})^2+\frac{\bar{r}^2}{4}\right\},\label{PolarPara}
\ee
where $\bar{r}=r/l_{\omega}$, with length unit $l_{\omega}=\sqrt{\hbar/2M\omega}$.

At the fermionic point, $\alpha=0$, there is no singular flux attached to anyons. Here, the exact single-particle solutions are $\phi_{n,m}$ with energy $E_{n,m}=(2n+m+1)\hbar\omega$,  where $n=0,1,\cdots$ is the Landau level index and $m=-n, -n+1, -n+2, \cdots$ is the $\hat{L}_{\rm z}$ component of angular momentum, in analogy with the Landau level solutions. 

When $\alpha\neq0$, the Kohn-Sham solutions $\psi_{n,m}$ and $\epsilon_{n,m}$ are modified but still carry the same quantum numbers. We note that, similarly to the familiar fermionic problem, the KS Hamiltonian in Eq.~\ref{PolarPara} is independent of $\omega$ and $M$ once we choose $\hbar\omega$ and $l_\omega$ as the units of energy and length. As a result, with these units, the KS orbitals are also independent of $M$ and $\omega$. This independence is consistent with the original anyon problem, where there is no gauge field to start with and the suitable length unit and energy unit are $l_{\omega}$ and $\hbar\omega$, respectively.

Therefore, within the KS formulation, the ground state of the anyon system evolves adiabatically as a function of the strength of the confining potential, as is also the case for non-interacting fermions in a parabolic potential. 

The same observation holds true in a hard wall potential:
\be
V_{\rm hw}(\vec{r})=0,\; r<R; V_{\rm hw}(\vec{r})=\infty, \; r\geq R.
\ee
in which Eq.~\ref{PolarPara} is rewritten as
\be
\mathcal H_{\rm KS}=\frac{\hbar^2}{2MR^2}\left\{  \left[-i\frac{1}{\bar{r}}\frac{\partial}{\partial\phi}+\frac{\alpha N(\bar{r})}{\bar{r}}\right]^2+(-i\frac{\partial}{\partial\bar{r}})^2\right\}+V_{\rm hw}(\bar{r})\;, \label{PolarHw}
\ee
where $\bar{r}=r/R$. Without an external magnetic field, the energy of the system is inversely proportional to the area of the hard wall potential well.  

We will be interested in the response to the application of an external magnetic field $B_{\rm ext}$, as a function of which the ground state is no longer guaranteed to evolve adiabatically. We first recall that for fermions, the external magnetic field perturbs the Hamiltonian with $\Delta H=\Delta H^{(1)}+\Delta H^{(2)}$ where $\Delta H^{(1)}\propto B_{\rm ext} \hat{L}_{\rm z}$ and $\Delta H^{(2)}\propto B^2_{\rm ext} r^2$. As a result we expect the total energy to change linearly with $B_{\rm ext}$ for small $B_{\rm ext}$ for ground states with $L_z\neq 0$. For filled shell configurations, which correspond to non-degenerate ground states, we have $L_{\rm z}=0$ and thus first order correction in $B_{\rm ext}$ vanishes leaving the energy to change quadratically in small $B_{\rm ext}$. For anyons, the situation is more complicated in that, first, $\Delta H\propto(\vec{p}+\frac{e}{c}\vec{A}_{\rm eff})\delta\vec{A}_{\rm ext}$ includes the coupling between $B_{\rm ext}$ and $B_{\rm eff}$, and second, $B_{\rm ext}$ changes the degeneracy of the effective Landau levels and thus may induce excitations across them in the KS solutions. 
The analysis for anyons is simplified in the limit of large $N$, because then the effective LL energy dominates over the confinement energy and the KS solutions describe a quantum Hall liquid. When $\alpha=1/n$ and $n$ is an integer, the ground state is non-degenerate and the lowest $n$ LLs are occupied. In that situation, there are level-crossing transitions as a function of $B_{\rm ext}$, which correspond to transfer of fermions between different effective LLs. Furthermore, we find that, in the thermodynamic limit, the energy increases when a small $B_{\rm ext}$ is applied, regardless of its direction; this has been interpreted as a Meissner-like behavior~\cite{Chen89}.

\section{Comparison with exact results}\label{Numerical190719}

\begin{figure}[t]
\includegraphics[width=3in, scale=1]{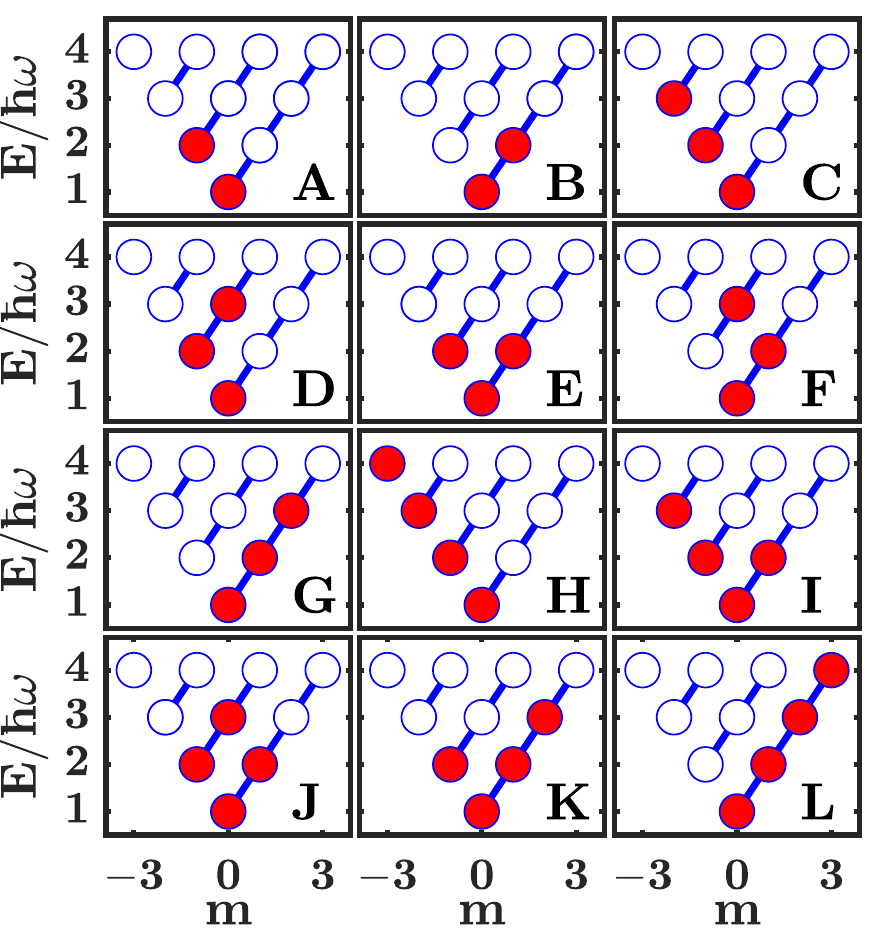}
\caption{This figure shows various low energy fermionic states (red circles indicate the occupied orbitals) for $N=2$, 3 and 4 particles. Each single-particle orbital is labeled by two quantum numbers $\{n,m\}$. The x-axis shows the angular momentum quantum number $m$. The circles joined by blue lines have the same ``Landau-level quantum number" $n$, with $n=0, 1, 2\ldots $ from right to left. The energy of an orbital with quantum numbers $\{n,m\}$ is given by $E_{n,m}=(2n+m+1)\hbar\omega$. The wave functions of these configurations, referred to as $\Phi_A, \cdots \Phi_L$, are used in trial wave functions. 
In our DFT formulation for anyons, we map the anyon problem into a problem of fermions carrying gauge fields; in constrained DFT we assume configurations given in this figure, and obtain the single-particle Kohn-Sham orbitals self-consistently.}\label{ConstrainOccu}
\end{figure}

\begin{figure*}[t]
\includegraphics[width=7in, scale=1]{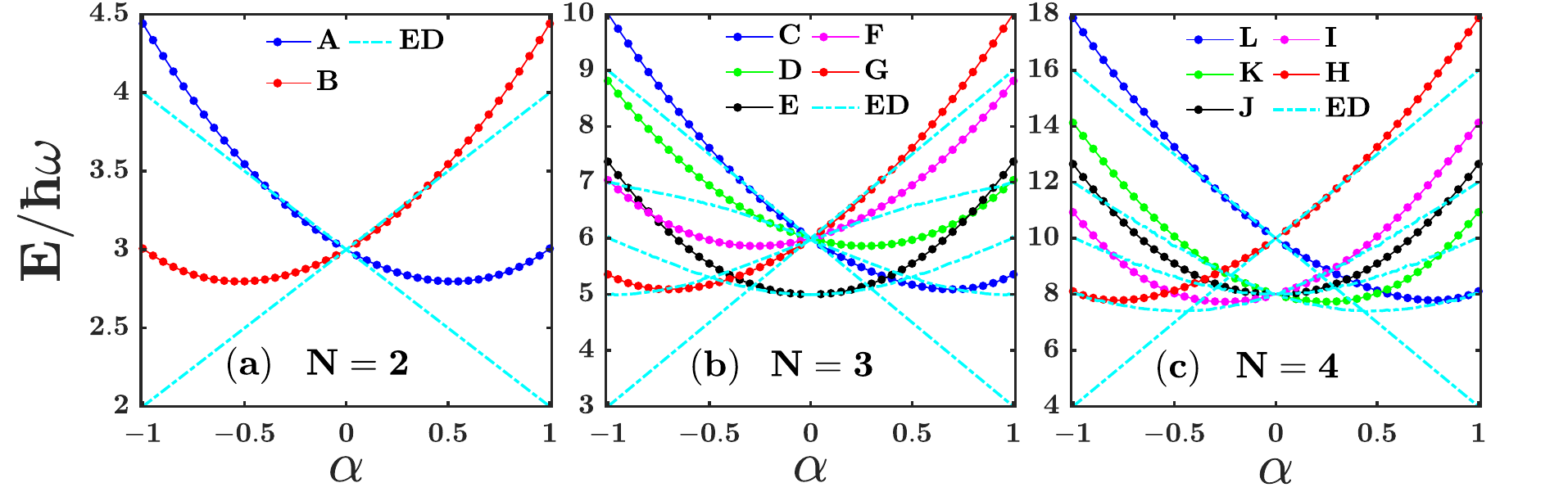}
\caption{Comparison between the energy obtained from constrained DFT calculation with exact results. The cyan lines are the low-energy spectra of $N=2, 3$ and $4$ anyons in a parabolic potential obtained using exact diagonalization (ED); these are taken from Sporre, Verbaarschot and Zahed \cite{Sporre91,Sporre92} and Murthy {\it et al.}\cite{Murthy91} The  circles are DFT results. The letters A to L correspond to the  occupation configurations shown in Fig.~\ref{ConstrainOccu}. }\label{ConstrainE}
\end{figure*}

In order to test the validity of the energy obtained by DFT, we compare it with the exact anyon spectrum in a parabolic potential. The exact spectrum of anyons in a parabolic potential is known for $N=2$~\cite{Wu84,Myrheim99}. 
Numerically obtained solutions for low energy eigenstates are known for systems of three and four anyons ~\cite{Sporre91,Murthy91,Sporre92,Sporre93}. In addition, two branches of analytical solutions, referred to as type I and type II, are known for arbitrary $N$, given by
\be
\psi_{\rm I}=\Delta_N^{1+\alpha}\exp(-\sum_{i}^N|z_i|^2/4l^2_{\omega})\label{ExactOne}
\ee 
\be
\psi_{\rm II}=(\Delta_N^*)^{1-\alpha}\exp(-\sum_{i}^N|z_i|^2/4l^2_{\omega})\label{ExactTwo}
\ee
with 
\be
\Delta_N=\prod_{1\leq i<j\leq N}(z_i-z_j).
\label{Delta}
\ee
For two particles, other solutions can be generated from them by a careful application of raising and lowering operators. All the known analytic solutions have energies that depend linearly on $\alpha$. The energies of $\psi_{\rm I}$  and $\psi_{\rm II}$ are $E_{\rm I}=[(N-1)\alpha+N+1]N/2$ and  $E_{\rm II}=[-(N-1)\alpha+N+1]N/2$ in units of $\hbar\omega$. We stress that in general, $\psi_{\rm I}$ and $\psi_{\rm II}$ are not ground states.

We obtain the spectra by implementing the so-called constrained DFT\cite{Kaduk12}. Specifically, we fix the orbitals as shown in Fig. \ref{ConstrainOccu} for $N=2$, 3 and 4, and obtain the self consistent solutions for the wave functions and the energies of these orbitals.  We employ the constrained DFT to enable us to follow the evolution of each fermionic Slater determinant state as a function of $\alpha$, and also to be able to treat both ground and excited states. The constrained DFT is to be contrasted with the standard DFT in which KS orbitals are occupied according to the Fermi-Dirac distribution in each DFT iteration; this would produce only the ground state.

The energies obtained from the constrained DFT are shown in Fig.~\ref{ConstrainE}, along with the exact energies known from earlier works\cite{Sporre91,Murthy91,Sporre92}. Near the fermionic point, the results of DFT show excellent agreement with the exact solutions. In particular, the DFT captures whether the behavior as a function of $\alpha$ is linear or quadratic, and in case of the former, also obtains the slopes to a high degree of accuracy [see Fig.\ref{ConstrainE}(a)]. Many features of the DFT results can be qualitatively explained from a first order perturbation theory in the effective magnetic field as follows. Flux attachment corresponds to a perturbation $\Delta H=[(\vec{p}+\vec{A}_{\rm eff})^2-\vec{p}^2]/2M$. To see the effect of the attached flux, we simplify the perturbation analysis by using the approximation $\vec{B}_{\rm eff}\propto  \alpha $ and thus $\vec{A}_{\rm eff}\propto  \alpha r$, i.e., we assume that the density is uniform in the region of interest. This allows us to rewrite $\Delta H=\Delta H^{(1)}+\Delta H^{(2)}$ where $\Delta H^{(1)}\propto \alpha \hat{L}_{\rm z}$ and $\Delta H^{(2)}\propto \alpha^2r^2$. This quadratic form of perturbation leads to an energy dispersion versus $\alpha$ in the shape of a parabola, with the location of its minimum being proportional to $L_{\rm z}$. This is consistent with the behavior seen in Fig.\ref{ConstrainE}(a-c). When $|\alpha|$ is small, $\Delta H^{(1)}$ dominates and the energy correction to first order is proportional to $\alpha L_{\rm z}$, as is seen from the DFT result around $\alpha=0$. These considerations also explain why, in the vicinity of $\alpha=0$, agreement over a larger range of $\alpha$ is seen on the side where the energy is increasing.

We also note that the exact spectrum shows a mirror symmetry relating $\alpha\rightarrow -\alpha$. In the DFT spectra, this symmetry is equivalent to invariance of the spectrum to a flip of effective magnetic field, i.e. to the replacement $z\rightarrow z^*$. As a result of this symmetry, the following features are observed:  Constrained states with ``symmetric" fermionic occupations [see Fig.~\ref{ConstrainOccu} E and J] also show a dispersion symmetric in $\alpha$ around $\alpha=0$. Also, two states, whose fermionic occupations are symmetric counterparts of one another (e.g. C and G, or I and K), have dispersion related by mirror symmetry around $\alpha=0$. 

\begin{figure}%[t]
\includegraphics[width=\columnwidth, scale=1]{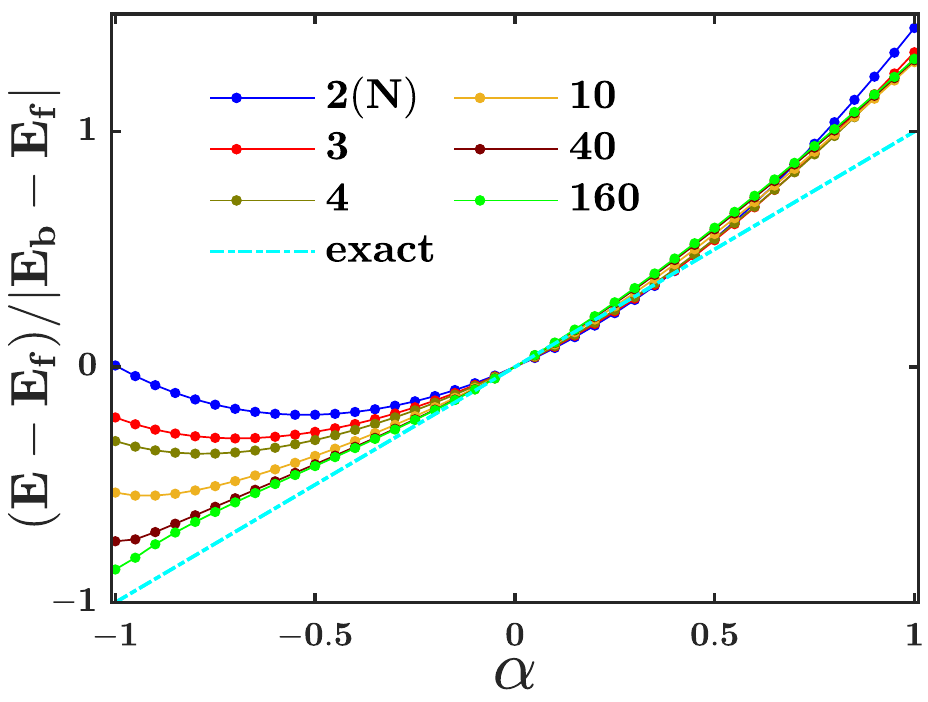}
\caption{Comparison of the energy obtained from constrained DFT with type $\rm I$ analytic solutions for several values of $N$. $E_f\equiv N(N+1)/2\hbar\omega$ is the energy at the fermionic point and $E_b\equiv N^2 \hbar\omega$ is the energy at the bosonic point. For type I analytic solutions, whose energies are linear in $\alpha$, $(E-E_f)/|E_b-E_f|$ is independent of $N$; this is shown by the cyan dashed line. The dots are obtained from the constrained DFT, where the occupied orbitals are $n=0, m=0, 1, \ldots, N-1$ for each particle number $N$. }\label{DFTedN}
\end{figure}

For larger $N$, the  exact ground state is  not available, so we compare energy from DFT with the available exact energy $E_{\rm I}$ of the type $\rm I$ solution. To compare the DFT energy with the exact energy, we plot in Fig.~\ref{DFTedN} the quantity $(E-E_f)/|E_b-E_f|$ where $E$ can be either the exact energy or the DFT energy. When $E$ is the exact energy, we have $(E_{\rm I}-E_f)/|E_b-E_f|=\alpha$ regardless of $N$, where $E_f=N(N+1)/2$ ($E_b=N^2$) is the exact energy at the fermionic (bosonic) end. For large $N$, we find that the DFT energy is again consistent with the exact energy near the fermionic point, where the total DFT energy shows a good linear relation versus $\alpha$ for a large range of $\alpha$. 

These comparisons show that our DFT formulation is qualitatively and semi-quantitatively valid close to the fermionic point. The deviations from exact results farther away from the fermionic point indicate the need for xc interaction. We have not attempted to add an xc interaction in this work, because our focus is largely on the topological properties. 

\section{Comparison with trial wave functions}\label{Trialsection}

\begin{figure}%[t]
\includegraphics[width=\columnwidth, scale=1]{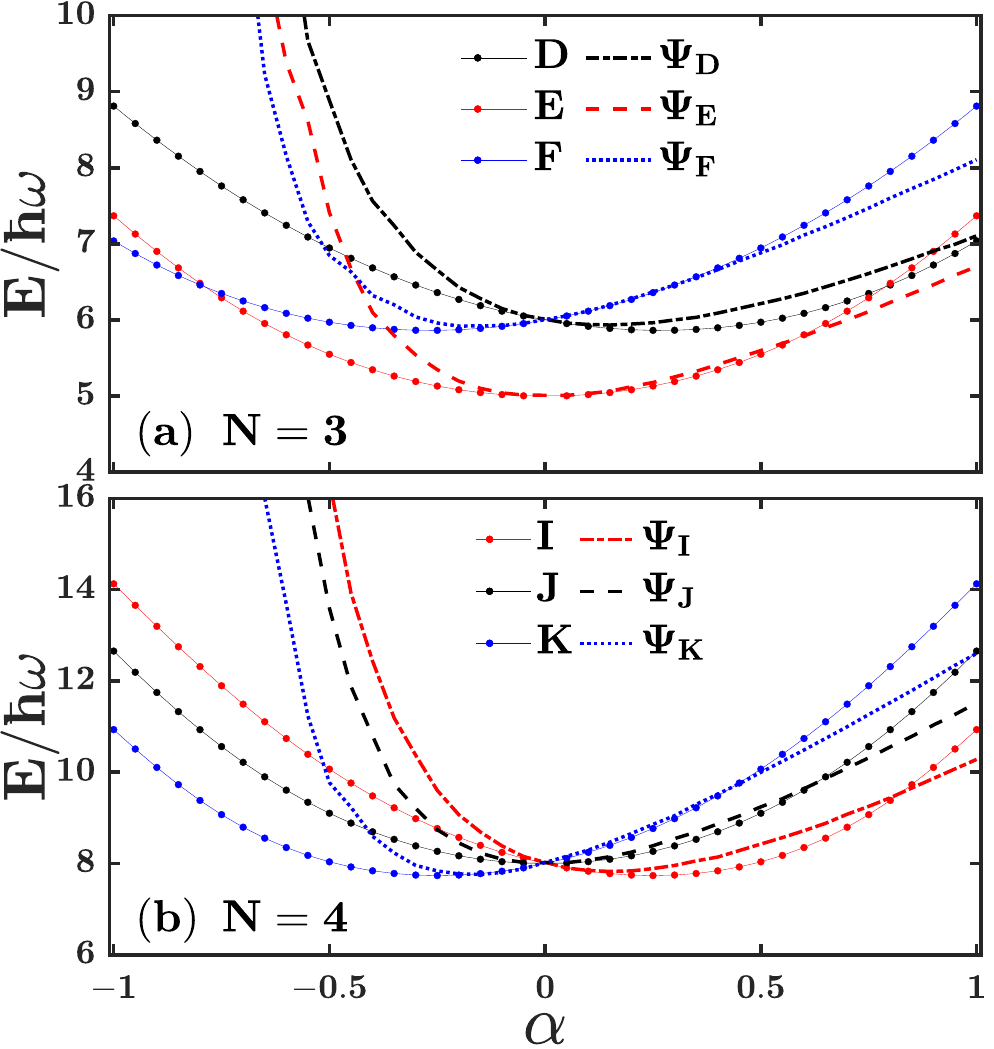}
\caption{Comparison of the energy obtained from constrained DFT (solid lines with dots) with the energy of the trial wave function in Eq.~\ref{TrialExpression} (dashed lines). The trial wave functions are denoted as $\Psi_X=[\Delta_N]^\alpha\Phi_X$ where X refers to the constrained occupation configurations shown in Fig.~\ref{ConstrainOccu}. For some cases, the trial wave function reduces to the exact analytic solution with energy linear in $\alpha$; comparison with DFT for these cases can be found in Fig.~\ref{ConstrainE}, for states labeled by A, B, C, G, H, L.}\label{ConstrainTrial}
\end{figure}

Trial wave functions for anyons can be constructed by close analogy to the wave functions for the composite fermion (CF) theory for 
FQHE. In the CF theory, one begins with IQHE states and then multiplies by the factor $[\Delta_N]^{2p}$ (see Eq.~\ref{Delta} for the definition of $\Delta_N$), where $p$ is an integer. This effectively attaches $2p$ flux quanta to each electron. This lends itself to a natural generalization to trial wave functions for anyons:
\be
\Psi_\alpha^{\rm trial}[\Phi]=[\Delta_N]^\alpha\Phi, \;\Psi_\alpha'^{\rm trial}[\Phi]=[\Delta^*_N]^{-\alpha}\Phi \label{TrialExpression}
\ee
where $\Phi$ denotes the Slater determinant wave functions of fermions, such as those shown in Fig.~\ref{ConstrainOccu}. Both of these wave functions manifestly describe anyons with statistics $1+\alpha$. The type I and type II wave functions in the earlier section above are also special cases of these trial wave functions. (As shown in Ref.\onlinecite{Hu19}, a DFT formalism similar to the one used in this paper provides a good representation of the composite fermion trial wave functions. We thus expect the DFT method we develop here for anyons  also to capture some features of the above trial wave functions.)

We calculate the total energy of $\Psi^{\rm trial}$ using the Monte Carlo sampling method. In each sampling step, the energy defined as $E=\frac{\mathcal{H}_A\Psi^{\rm trial}_{MC}}{\Psi^{\rm trial}_{MC}}$ is calculated by taking derivatives using the finite difference method. The total energy is averaged over 80000 sampling points. These energies are shown in Fig.\ref{ConstrainTrial} for systems with $N=3$ and $4$. This figure also shows the DFT energies (same as those in the previous figures). 

The energies of the trial wave functions $\Psi_\alpha'^{\rm trial}[\Phi]$ are not given explicitly, because they are the same as the energies of $\Psi_{-\alpha}^{\rm trial}[\Phi^*]$. This follows from the relation
\be
\Psi_\alpha'^{\rm trial}[\Phi]=\{[\Delta_N]^{-\alpha}\Phi^*\}^*=\{\Psi_{-\alpha}^{\rm trial}[\Phi^*]\}^*\;.
\ee
For example, the energy of $\Psi_\alpha'^{\rm trial}[\Phi_{\rm I}]$ is the same as that of $\Psi_{-\alpha}^{\rm trial}[\Phi_{\rm K}]$, because $\Phi_{\rm K}=\Phi_{\rm I}^*$. (States in Fig.~\ref{ConstrainE} that are related by mirror transformation about the vertical are complex conjugates of one another, such as A and B, or C and G. Certain states, such as E or J are self conjugate.) 

We see that the energies of $\Psi^{\rm trial}$ are consistent with the DFT energies in a large range of $\alpha$ near the fermionic point for $\alpha>0$. The larger increase in the trial wave function energy for $\alpha<0$ is due to the fact that $\Delta_N^{\alpha}$ contributes a large kinetic energy when $\alpha<0$. For $\alpha<0$, the DFT results better match the energy of the trial wave function $\Psi_\alpha'^{\rm trial}[\Phi]$. For example, the trial wave function energy shown by the blue dashed line for $\alpha>0$ in Fig.\ref{ConstrainTrial}(b)  nicely matches, upon flipping to $\alpha<0$, with the DFT energy of red solid line for $\alpha<0$ in the same figure.

For the state J, which is invariant under complex conjugation ($\Phi_{\rm J}=\Phi_{\rm J}^*$), the DFT energy is invariant under $\alpha\rightarrow -\alpha$. However, while $\Psi_\alpha[\Phi_J]$ gives a better energy for $\alpha>0$, $\Psi'_\alpha[\Phi_J]$  gives a better energy for $\alpha<0$ (the energy of $\Psi'_\alpha[\Phi_J]$ is identical to the energy of $\Psi_{-\alpha}[\Phi_J]$).

In conclusion, we find that the results of {\it constrained} DFT are consistent with the corresponding trial wave functions. We have shown this for both small systems of $N\leq4$ and large systems, although for the latter we have only made comparisons for cases where the trial wave function reduces to Eq.~\ref{ExactOne} or ~\ref{ExactTwo}. This implies that the bare DFT model under consideration is representative of the trial states we propose. This is expected from the facts that a similar DFT method in FQHE also gives results consistent with the trial wave functions of composite fermions, and that the trial states of anyons have structure similar to that of the composite fermion wave functions. 

With that established, the full power of DFT, which takes it beyond trial wave functions, is that it enables determination of the ground state density profile and energy in arbitrary external potentials. It also allows a treatment of large systems with much reduced computational cost. In principle, the DFT also can be improved with better choices of the xc potential (which has been neglected in this article).

We note that one can improve the trial wave functions by introducing a ``regulating parameter" introduced by Lundholm~\cite{Lundholm17,Lundholm13}. We give an example in the Appendix~\ref{regularize}. However, a systematic improvement of the trial states is beyond the scope of this article, which is devoted to the KS-DFT treatment of anyons.

\section{Large number of anyons in a confining potential}\label{Largesection}

In this section, we focus on the properties of a large number of anyons. We consider anyons 
with uniform density, as well as anyons trapped in a parabolic or hard wall  confinement.  We specialize to the statistical parameter $\alpha=1/n$, with $n$ being an integer. Anyons with these values of $\alpha$ fill $n$ Landau levels in the mean field theory and are expected to exhibit superconductivity. 

\subsection{Anyons with uniform density}

Let us first consider a system of anyons with uniform density $\rho$, which occurs when the external potential is a constant. Here, the KS equation describes particles moving in a uniform magnetic field
\be
B=B_{\rm ext}+B_{\rm eff}\;,\label{totalBscalar}
\ee
with the effective statistical magnetic field being
\be
B_{\rm eff}=\alpha\rho\phi_0\;,\label{Beffscalar}
\ee
and the KS orbitals form Landau levels in magnetic field $B$. The ``effective" filling factor $\nu^*$ of the KS system depends on both the statistical parameter and the external magnetic field:
\be
\nu^*=\rho\phi_0/B\;,\label{nuStarDef} 
\ee
while the ``real" filling factor $\nu$ of the system depends only on the external magnetic field:
\be
\nu=\rho\phi_0/B_{\rm ext}\;.\label{nuDef} 
\ee
When $B_{\rm ext}$ is zero, we have $\nu=\infty$ whereas $\nu^*=1/\alpha$, i.e. the lowest $\nu^*=\rho\phi_0/B_{\rm eff}$ Landau levels are occupied in the ground state. When $B_{\rm ext}$ is not zero, the general relation between $\nu$ and $\nu^*$ is:
\be
\nu=\frac{\nu^*}{1-\alpha \nu^*}\;,\label{RealFilling}
\ee
which can be obtained through the relations in Eqs.~\ref{totalBscalar}-\ref{nuDef}. Integer quantum Hall effect of anyons is obtained when $\nu^*=n^*$ is an integer, which corresponds to real filling factors given by
\be 
\nu={n^*\over 1-\alpha n^*}.\label{RealFillingnu*}
\ee

Of interest below is the behavior of the total energy of the system as a function of $\nu$ (or $B_{\rm ext}$)   in the vicinity of $\nu^*=n^*$. To see the physics for a uniform system, let us suppose that an IQHE of anyons with filling factor $n^*$ occurs at a certain value of $B_{\rm ext}$.  Let us examine how the energy of anyons changes when we slightly vary the external magnetic field, while holding the density fixed. When we apply a tiny additional external magnetic $\delta \vec{B}$, with $\delta B>0$ $(\delta B<0)$ when it is (anti) parallel to $\vec{B}$, the new filling factor $\nu^*=n^*-\delta n^*$ is given by
\be
(B+ \delta B)(n^*- \delta n^*)=B n^*\;,\label{densityConserve}
\ee
where we have assumed $\rho$ is fixed. The total energy per unit area is given by
\begin{eqnarray}
E&=&\frac{\hbar e(B+\delta B)^2}{Mc\phi_0}\left\{\sum_{i=0}^{n^*-1}(i+\frac{1}{2})-(n^*\mp\frac{1}{2})\delta n^*\right\} \nonumber \\ 
&=&\frac{\pi\hbar^2n^{*2}}{M\phi_0^2}\left\{B^2\pm\frac{1}{n^*}B\delta B-(1\mp\frac{1}{n^*})(\delta B)^2 \right\},\label{Meissner}
\end{eqnarray}
where the upper (lower) sign choice is picked when $\delta B>0$ $(\delta B<0)$. The important point here is that, to linear order in $\delta B$, the energy increases independently of the sign of $\delta B$. This behavior was obtained in Ref.~\onlinecite{Chen89} 
for the special case of anyons with $\alpha=1/n$ and $B_{\rm ext}\approx 0$, and interpreted in terms of a Meissner effect and, thus, as a possible signature of anyon superconductivity\cite{Fetter89,Chen89}. For larger external magnetic fields,  integer values of $n^*$ produce integer quantum Hall effect of anyons.

\subsection{Anyons in a parabolic potential}\label{sectionPara}

\subsubsection{Density and energy in zero magnetic field}

Some examples of self-consistent density distributions obtained from DFT for anyons in a parabolic potential $(1/2) M\omega^2r^2$ are shown in Fig.~\ref{Potentialparabolic}(a). It is seen that the density of anyons decays quadratically from the center, approximately in the same fashion as that of fermions, regardless of the statistical parameter $\alpha$. That remains true for even larger systems consisting of several thousand particles that we have studied. The total energy of anyons versus particle number $N$ is shown in Fig.~\ref{Potentialparabolic}(b). We find that the total energy also scales as $N^{3/2}$ similar to fermions, independent of the value of $\alpha$. 

To gain insight into these results from our DFT method, in the following, we look at the problem in a semiclassical Thomas-Fermi approximation\cite{Chitra92,Li92}, which is expected to be valid for noninteracting fermions when the particle density is large enough that $\rho(\vec{r}) l^2_{\omega}\gg (r/l_{\omega})^{2/3}$ in two dimensions\cite{Butts97}.  According to the Thomas-Fermi approximation, at each point $\vec{r}$, we write the total energy as (see Appendix~\ref{classicalDensity} for details)
\be
\frac{[\hbar \vec{k}_{\rm F}(\vec{r})+\frac{e}{c}\vec{A}(\vec{r})]^2}{2M}+\frac{1}{2}M\omega^2r^2=E_{\rm F}\;,\label{AnyonClassical}
\ee
where $\vec{A}(\vec{r})$ includes both the external vector potential and the effective vector potential due to the particle statistics. 
When $\vec{A}(\vec{r})=0$, Eq.~\ref{AnyonClassical} describes fermions in zero applied magnetic field and the solutions of the density distribution $\rho_{\rm TF}$ and energy $E_{\rm TF}$ are given in Appendix~\ref{classicalDensity}. For this case, as shown in  Appendix~\ref{classicalDensity}, the density at the center is $\rho_{\rm TF}(r=0)=\frac{\sqrt{2N}}{2}$ in units of $(2\pi l^2_{\omega})^{-1}$;  $\rho_{\rm TF}(\vec{r})$ falls off quadratically from the center, and vanishes for $r>R_{\rm F}$ where $R_{\rm F}=2(2N)^{1/4}l_{\omega}$ defines the edge. For anyons, we have $\vec{A}(\vec{r})\neq0$. However the vector potential only shifts the local wave vector by a constant vector $-\frac{e}{c}\vec{A}(\vec{r})$ in the momentum space. As a result, the vector potential moves the origin of the ``local" Fermi circle but does not affect the volume enclosed by the Fermi surface, i.e., the local density; the local average kinetic energy and potential energy also do not change. In other words, the mean-field flux attachment to fermions has no effect in the semiclassical Thomas-Fermi approximation. This gives some insight into why the density and energy are rather insensitive to the statistics parameter.

\subsubsection{Response to an external magnetic field: ``Meissner effect"}
We next study the energy  response of anyons to an external magnetic field.
It is not immediately obvious how the behavior described previously in Eq.~\ref{Meissner} is modified when the density of anyons is not uniform, for example, when they are in a confining potential. One might think that if the density is uniform on the scale of the effective magnetic length, then the above argument applies locally. However, as our explicit calculation shows, the situation is much subtler than expected.

\begin{figure}
\includegraphics[width=\columnwidth]{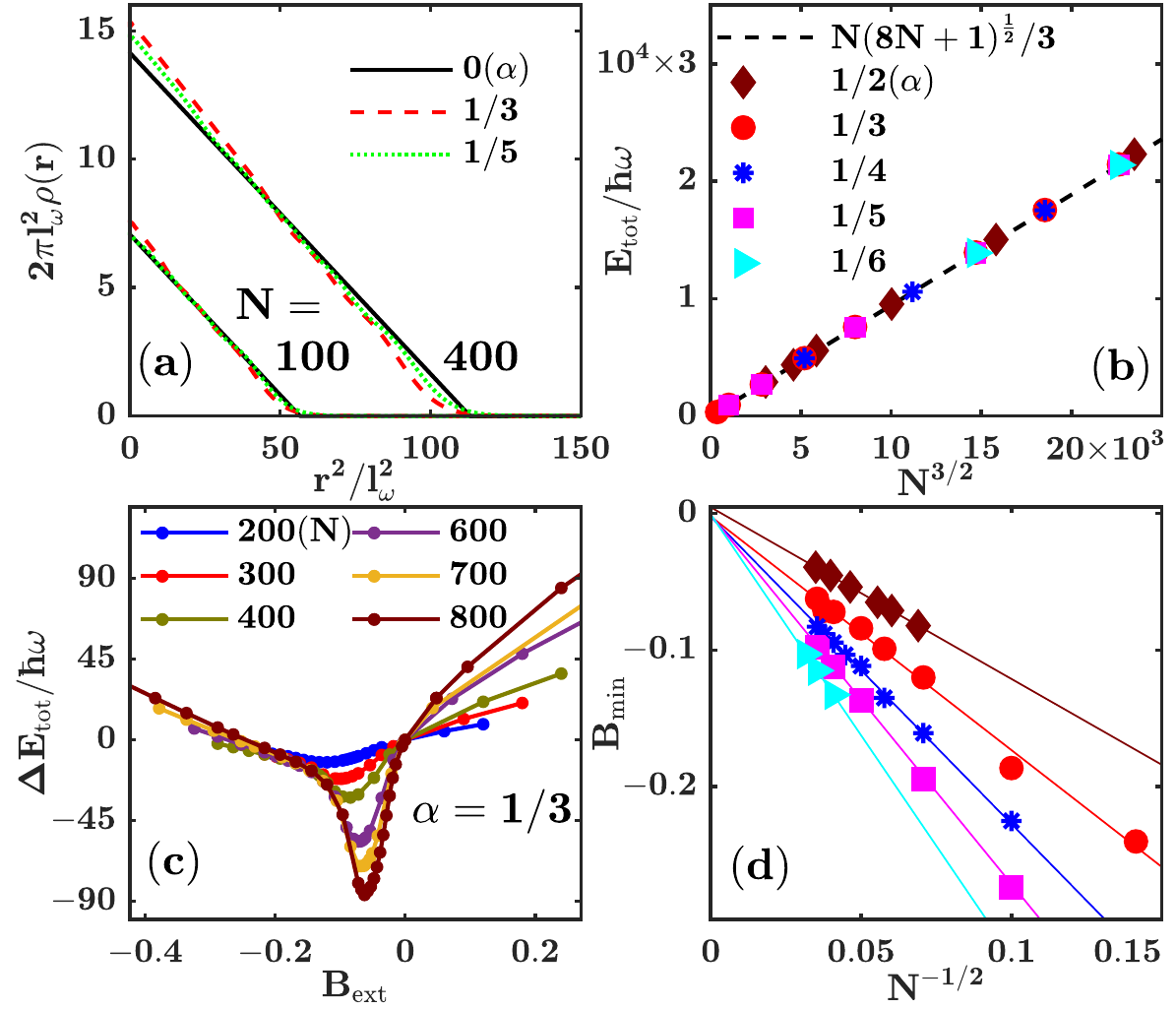}
\caption{This figure shows results for anyons in a parabolic potential $(1/2)M\omega^2 r^2$. (a) The density distribution of anyons with $\alpha=1/3$ (red) and $\alpha=1/5$ (green). The black line represents an approximate density profile for fermions ($\alpha=0$) obtained using the Thomas-Fermi approximation, which produces a quadratic decay for the density profile. (b) Total energy of anyons with $\alpha=1/2, 1/3, 1/4, 1/5, 1/6$ versus the particle number. The dashed line is given by $N(8N+1)^{1/2}/3$, which is the total energy for fermions with filled shell configurations. (c) Increase of total energy as a function of an external magnetic field, defined as $\Delta E_{\rm tot}(B_{\rm ext})=E_{\rm tot}(B_{\rm ext})-E_{\rm tot}(B_{\rm ext}=0)$. Its minimum occurs at magnetic field $B_{\rm min} $. (d) Scaling of $B_{\rm min} $ versus $1/\sqrt{N}$. $B_{\rm min} $ appears to extrapolate to zero in the thermodynamic limit. The symbols in panel (d) have the same definitions as in panel (b). $B_{\rm ext}$ and $B_{\rm min}$ are quoted in units of $Mc\omega/e$. For all calculations, we have assumed a negligibly small temperature $k_{\rm B}\tau=0.005\hbar\omega$ to facilitate convergence. }
\label{Potentialparabolic}
\end{figure}

We calculate the change of total energy by applying a small external magnetic field $B_{\rm ext}$ in a parabolic potential using the DFT method. The results for different numbers of anyons with $\alpha=1/3$ are shown in Fig.~\ref{Potentialparabolic}(c). When $B_{\rm ext}$ is parallel to $B_{\rm eff}$, the total energy increases with the increasing of $B_{\rm ext}$, consistent with Eq.~\ref{Meissner}. However, when $B_{\rm ext}$ is anti-parallel to $B_{\rm eff}$, the energy decreases, in contrast to the behavior predicted by Eq.~\ref{Meissner}.  This inconsistency results from the  finite-size of the system under consideration, because of which two of the assumptions made above are not valid. First, in our DFT calculation, we notice that upon the application of $B_{\rm ext}$ in the anti-parallel direction, some of the density from the bulk is pumped into the edge; therefore, the assumption of the conservation of local density assumed in Eq.~\ref{densityConserve} is no longer valid. Second, while the derivation of Meissner-like effect in Eq.~\ref{Meissner} requires that the filling factor is precisely $n$ when no external magnetic field is applied, in our problem with parabolic confinement, the filling factor in the interior is smaller than $n$.

Nonetheless, the energy shows a minimum at a slightly negative $B_{\rm ext}$, which we label  $B_{\rm min}$. In Fig.~\ref{Potentialparabolic}(d) we find that the value of $B_{\rm min}$ approaches zero in the thermodynamic limit. In other words, the behavior predicted by mean-field theory for a uniform density system is recovered in the thermodynamic limit. In the next subsection, we provide a semi-quantitative discussion of this behavior using a modified mean field theory that takes into account the effect of parabolic confinement. 

\subsubsection{Mean field approximation in a parabolic confinement}

The parabolic confinement leads to modifications in the single particle wave function as well as in the definition of the filling factor. 
We assume that the effective magnetic field $B_{\rm eff}$, and thus also the density, is uniform in the region of interest. This approximation is valid at least in the bulk surrounding the center where $B_{\rm eff}\propto \alpha\frac{\sqrt{2N}}{2}$ is dominant and the parabolic potential can be considered as smooth on the scale of the magnetic length. The single particle orbitals in a parabolic confinement subjected to a uniform magnetic field are given by the Fock-Darwin levels, with a dispersion $E_{n, m}=\hbar\Omega(n+1/2)+\hbar(\Omega-\omega_{\rm eff})(m-n)/2$, where $\Omega=\sqrt{\omega^2_{\rm eff}+4\omega^2}$ is the Fock-Darwin cyclotron gap and $\omega_{\rm eff}=\frac{eB_{\rm eff}}{Mc}$. In the Fock-Darwin solution, the Fock-Darwin length unit is $l_{\Omega}=\sqrt{\hbar/M\Omega}$, which replaces the magnetic length. The effective filling factor is 
\be
n^*=2\pi l^2_{\Omega}\rho=\frac{\rho\phi_0}{\sqrt{(B_{\rm eff}+B_{\rm ext})^2+(\frac{2Mc\omega}{e})^2} }.\label{argument}
\ee
Notice that if we have $B_{\rm ext}=0$ and $\omega=0$, then it follows that $n^*=1/\alpha=n$. However, for $B_{\rm ext}=0$ and $\omega\neq0$, we have $n^*<1/\alpha=n$. 
The energy minimum occurs when the bulk is integrally occupied, i.e., $n^*=n$. This, in turn, occurs when we have $\sqrt{(B_{\rm eff}+B_{\rm ext})^2+(\frac{2Mc\omega}{e})^2}=B_{\rm eff}$, the solution of which for $B_{\rm ext}$ yields 
\be
B_{\rm min}\approx-\frac{2}{B_{\rm eff}}(Mc\omega/e)^2\propto \frac{1}{\alpha\sqrt{N}}.
\label{scaling}
\ee  This scaling relation $B_{\rm min}\propto 1/\sqrt{N}$ is numerically confirmed for different statistical parameters $\alpha$, as shown in Fig.~\ref{Potentialparabolic}(d). 
If we take $B_{\rm eff}$ in Eq.~\ref{scaling} to be the $B_{\rm eff}(r=0)$ of our DFT calculation, then the resulting value of $B_{\rm min}$ from Eq.~\ref{scaling} is approximately a factor of two smaller than the $B_{\rm min}$ obtained from our DFT formulation. This discrepancy is not significant, given that the derivation of Eq.~\ref{scaling} assumes a constant density, which is not the case for the DFT solution.

The above mean field analysis also explains the insensitivity of the radius $R_F$ of the density distribution and of the total energy to the value of $\alpha=1/n$. This can be seen as follows. Near the edge, the harmonic potential becomes dominant and traps $N\alpha$ particles in each Landau level due to its sharp confinement. This gives the radius of the anyon distribution as $\sqrt{N\alpha}l_{B_{\rm eff}}$, which is seen to be independent of $\alpha$ in view of $l_{B_{\rm eff}}=\sqrt{ \frac{\hbar c}{e |B_{\rm eff}|} }\propto 1/\sqrt{\alpha}$. In addition, the average energy per particle in the bulk is $E_{\rm tot}/N=\hbar\omega_{\rm eff}n/2\propto \sqrt{N}$, independent of $\alpha=1/n$, thus also leading to the observed scaling behavior of $E_{\rm tot}\propto N^{\frac{3}{2}}$. 

Incidentally, we can also obtain $B_{\rm eff}$ from the mean field theory, by using the equation $\sqrt{N\alpha}l_{B_{\rm eff}}=R_{\rm F}$, which implies $B_{\rm eff}=\alpha\phi_0 N/\pi R^2_{\rm F}$. This is exactly the flux density attached to the average particle density inside $R_{\rm F}$.

The deviation of $B_{\rm min}$ from zero is also seen in other confinement potentials like the hard wall potential (see Sec.~\ref{HWsection}) and a cubic potential $V_{\rm ext}\propto r^3$ (results not shown). In all the  cases that we tested, the magnitude of $B_{\rm min}$ decreases with increasing $N$. We expect that to be a generic feature, since when $N$ increases, $B_{\rm eff}$ increases and the physics is dominated by Landau levels rather than the confining potential. Consequently, this creates a large uniform bulk and Eq.~\ref{densityConserve} is restored when the edge effects are negligible. We point out the calculations presented in this section assume a finite temperature on the order of $k_{\rm B}\tau=0.005\hbar\omega$ for the purpose of convergence. The temperature is much smaller than the effective cyclotron energy of the anyons, and we have found that the results do not change appreciably for a temperature range of $k_{\rm B}\tau=0.05-0.005\hbar\omega$. Hence, our results can be taken as representing the zero temperature limit.

\begin{figure}[t]
\includegraphics[width=\columnwidth]{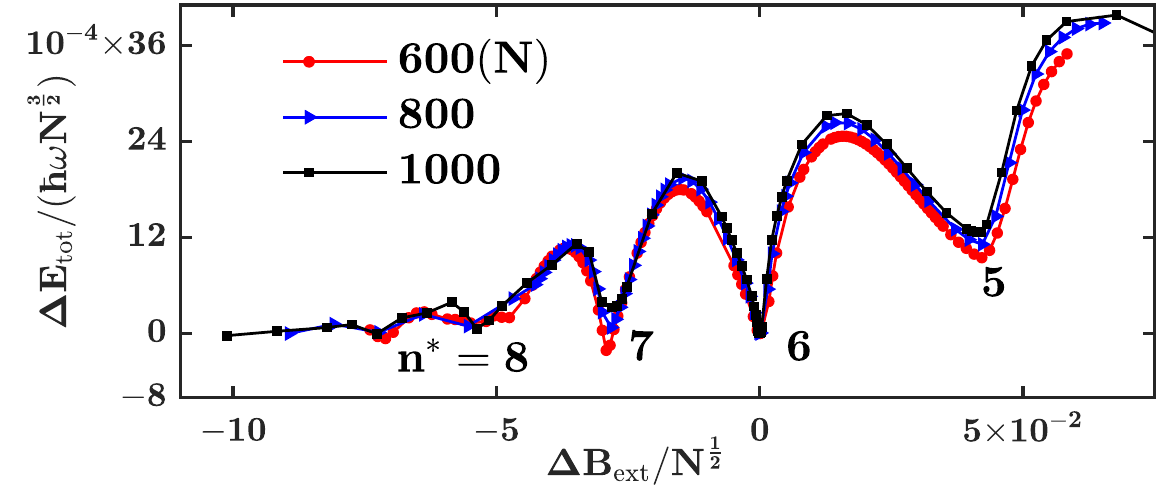}
\caption{ Change of total energy as a function of the magnetic field for anyons with $\alpha=1/6$ at temperature $k_{\rm B}\tau=0.005\hbar\omega$. The energy minima are labeled by the effective filling factor $n^*$ around which they occur. For each $N$, $\Delta E_{\rm tot}$ and $\Delta B_{\rm ext}$ are measured with reference to the values of $E_{\rm tot}$ and $B_{\rm ext}$ at the minimum labeled by $n^*=6$. The external magnetic field is quoted in units of $Mc\omega/e$.}\label{ManyNstarHWFig}
\end{figure}

\subsubsection{Quantum Hall effect}

In the above, we have studied the response of anyons to a small $B_{\rm ext}$. We now consider anyons at arbitrary $B_{\rm ext}$. For a uniform density, anyons show integer quantum Hall effect when $\rho\phi_0/B=n^*$ where $n^*$ is an integer and $B=B_{\rm ext}+\alpha \rho\phi_0$. We see below how this behavior comes about in our DFT calculations for anyons with nonuniform densities.

According to Eqs.~\ref{densityConserve} and \ref{Meissner}, a local energy minimum versus $B_{\rm ext}$ occurs whenever an integer number, labeled $n^*$, of Landau levels are filled in the bulk. These energy minima are indeed observed in our DFT calculation in Fig.~\ref{ManyNstarHWFig}, which shows the energy of the system as a function of $B_{\rm ext}$ for $\alpha=1/6$ with different particle numbers. The successive minima correspond to IQHE of anyons.

Determining the filling factor is complicated because the density is nonuniform and also, the density at the center grows as $\sqrt{N}$ in the parabolic geometry. For that reason, we find it convenient to plot the energy as a function of $\Delta B_{\rm ext}/\sqrt{N}$, where $\Delta B_{\rm ext}=B_{\rm ext}-B_{\rm min}$ and $B_{\rm min}$ is the position of the minimum close to $B_{\rm ext}=0$. The scaling with $\sqrt{N}$ reflects the fact that the density grows as $\sqrt{N}$.  The energy $\Delta E_{\rm tot}$ is measured relative to the energy at $B_{\rm min}$. We find that $\Delta E_{\rm tot}$ scales as $N^{3/2}$, just as the ground state energy (see Appendix \ref{classicalDensity}).

We note that the scaling behavior is expected to be valid only in the thermodynamic limit, which is why there is slight deviation between the curves for different $N$ in Fig.~\ref{ManyNstarHWFig}. 
As we see in Sec.~\ref{HWsection}, the thermodynamic limit is obtained more readily in the hard wall configuration, where the edge effects are relatively small.

\subsection{Anyons in a hard wall potential}\label{HWsection}

In the following we study the properties of anyons with $\alpha=1/n$ in a hard wall potential of radius $R$. We look at the total energy and its response to an external magnetic field, as well as the quantum Hall effect of anyons. The scaling behavior of anyons in this confinement is different from that in a parabolic potential. 

We first show that the total energy of the system scales as $E_{\rm total}\sim N^2$, for $B_{\rm ext}=0$. This behavior can be derived using the mean field approximation which works well for the hard wall potential. On a mean field level, the effective magnetic field is 
\be 
B_{\rm eff}=\alpha NB_1,\;\; B_{\rm 1}\equiv \phi_0/\pi R^2\;, \label{B1Definition}
\ee
with the cyclotron gap given by $\alpha N\hbar\omega_{1}$, with $\omega_{\rm 1}=eB_{\rm 1}/Mc$. The field $B_1$ is the magnetic field generated by a single uniform flux inside the potential well. We often use $l_{B_{1}}=\sqrt{ \frac{\hbar c}{e |B_{1}|} }$ as the length unit, because the radius of the potential well $R/l_{B_1}=\sqrt{2}$ becomes a constant in this unit. 

\begin{figure}[t]
\includegraphics[width=\columnwidth]{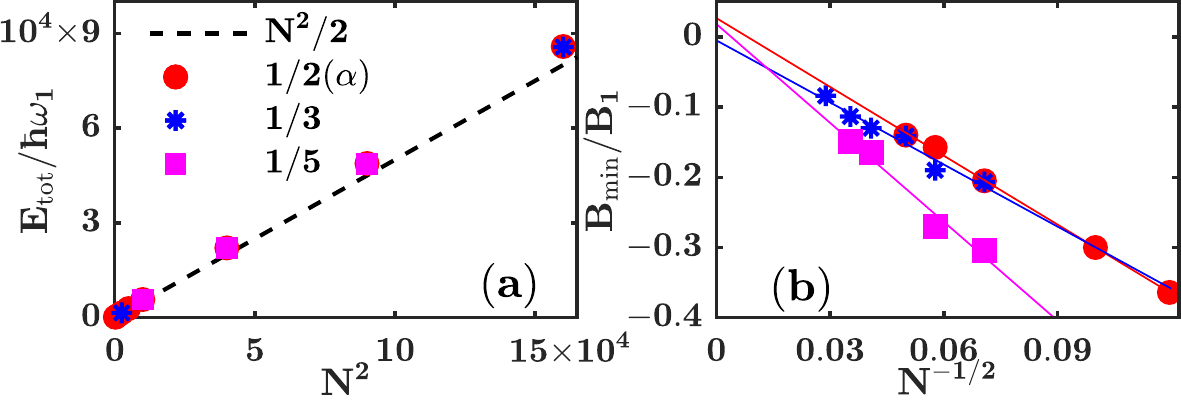}
\caption{ This figure shows results for anyons in a hard wall potential. 
We quote the energy in units of $\hbar \omega_{1}=\hbar \frac{eB_{1}}{Mc}$, which is the cyclotron gap at a magnetic field $B_1=\frac{\phi_0}{\pi R^2}$ generated by a single flux quantum uniformly spread out in the potential well of radius $R$; the effective cyclotron energy of the anyon system at $B_{\rm ext}=0$ is then given by $\alpha N \hbar \omega_{1}$. (a) Total energy of anyons as a function of the particle number for $\alpha=1/2, 1/3, 1/5$. The dashed line shows the mean-field approximation of $E_{\rm tot}= \hbar\omega_{1}N^2/2$. The deviation between the DFT energy and the mean-field energy arises from the non-uniformity of the density due to the confinement. (b) $B_{\rm min}$ is shown as a function of $1/\sqrt{N}$, where $B_{\rm min}$ is the value of the external magnetic field at which the total energy has a minimum in the vicinity of $B_{\rm ext}=0$. The symbols in panel (b) have the same definitions as those in panel (a). All calculations are performed at a negligibly small finite temperature $k_{\rm B}\tau=\hbar\omega_{1}$.}\label{PotentialWellB}
\end{figure}

When $\alpha=1/n$, the energy per particle is $\alpha N\hbar\omega_{1}n/2$ and the total energy is $E_{\rm tot}= \hbar\omega_{1}N^2/2$, independent of $n$, as can be seen in Fig.~\ref{PotentialWellB}(a). In the hard wall potential, it is convenient to use $\omega_{\rm 1}$ as the unit of energy, such that the results of energy reported in this paper are independent of the radius $R$ of the potential well when $B_{\rm ext}=0$. 

This behavior of $E_{\rm tot}$ is also related to a result from the magnetic-Thomas-Fermi theory,\cite{Lieb95}, which predicts that the total energy of electrons scales as $E_{\rm tot}\propto N^2$ in the large $N$ limit if both the magnetic field $B$ and the external confinement $V_{\rm ext}$ are proportional to the electron number $N$. The role of the second condition of $V_{\rm ext} \propto N$ in the work of Ref.~\onlinecite{Lieb95} is essentially to guarantee that the area of the density distribution is fixed, i.e., to approach a high density limit when $N\rightarrow \infty$.  
In the hard wall potential, both these conditions are satisfied: the effective magnetic field $B_{\rm eff}$ now plays the role of the external magnetic field, which is proportional to $N$; additionally, the area is fixed by the hard wall confinement. 

We now consider the energy response to a tiny magnetic field $B_{\rm ext}$. An energy minimum similar to that in the parabolic potential occurs at $B_{\rm ext}=B_{\rm min}$. In comparison with the parabolic potential, the density is smoother in a hard wall potential and the deviation of $B_{\rm min}$ from zero is a much smaller. It is expected that $B_{\rm min}$ will approach zero in the thermodynamic limit, for the same reasons as those mentioned in the context of parabolic potential. From the extrapolation of $B_{\rm min}$ versus $1/\sqrt{N}$ shown in Fig.~\ref{PotentialWellB}(b), it is seen that $B_{\rm min}$ indeed approaches zero within numerical uncertainty; our calculation indicates that extrapolation of $B_{\rm min}$ may require terms with higher orders of $1/\sqrt{N}$. 

\begin{figure}[t]
\includegraphics[width=\columnwidth]{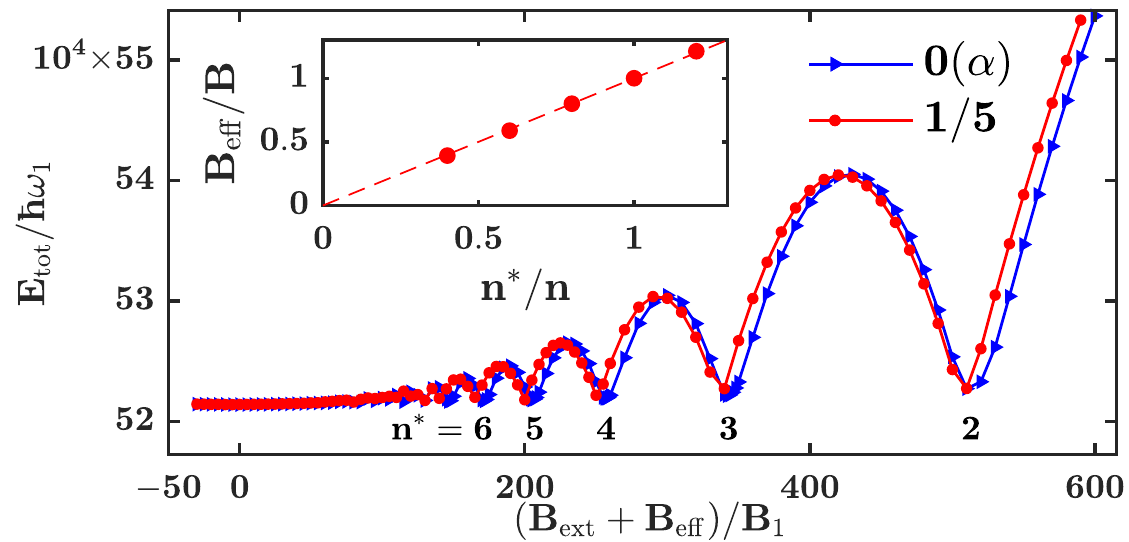}
\caption{Energy of anyons with $\alpha=1/5$ as a function of the total magnetic field $B=B_{\rm ext}+B_{\rm eff}$. For comparison, the energy of fermions ($\alpha=0$) is also shown. Here the effective magnetic field is chosen to be $B_{\rm eff}=\alpha N\phi_0/(\pi R^2)$, i.e., the mean field value generated by the flux attached to anyons. The inset shows the total magnetic field $B=B_{\rm ext}+B_{\rm eff}$ at the energy minima labeled by $n^*=2, 3, 4, 5, 6$ in the main panel. These fall on the dashed line that is a plot of the relation $Bn^*=B_{\rm eff}n$. The system contains $N=1000$ particles and the temperature is $k_{\rm B}\tau=10\hbar\omega_{1}$. }\label{SdHHW}
\end{figure}

Finally we look at the quantum Hall effect of anyons in the hard wall potential. As explained earlier, the energy shows a minimum for each IQHE state. 
We show in Fig.~\ref{SdHHW} the energy as a function of $B_{\rm ext}$ for fermions, as well as for anyons with $\alpha=1/5$. The minima correspond to the integer values of $n^*$ shown on the figure, which is the effective filling factor $\nu^*$ in the bulk; there are slight corrections due to the boundary which will become negligible in the thermodynamic limit. The values of $B=B_{\rm ext}+B_{\rm eff}$ where the minima occur for a uniform density system are given by $Bn^*=B_{\rm eff}n$, as seen from Eq.~\ref{densityConserve}. 
The inset demonstrates that the values of $B_{\rm ext}$ at the minima are consistent with this relation. Of course, when plotted as a function of $1/(B_{\rm ext}+B_{\rm eff})$, the anyon system exhibits Shubnikov-de Haas oscillations. 

The fact that the energy curves for fermions and anyons are almost the same when we plot them as a function of $B=B_{\rm ext}+B_{\rm eff}$ demonstrates that the statistical field behaves similarly to the real magnetic field. This is a feature of the mean field theory for uniform density and is also consistent with exact solutions~\cite{Canright89}. 

\section{Topological properties of IQHE}\label{topogicalQH}

\begin{figure}[t]
\includegraphics[width=\columnwidth]{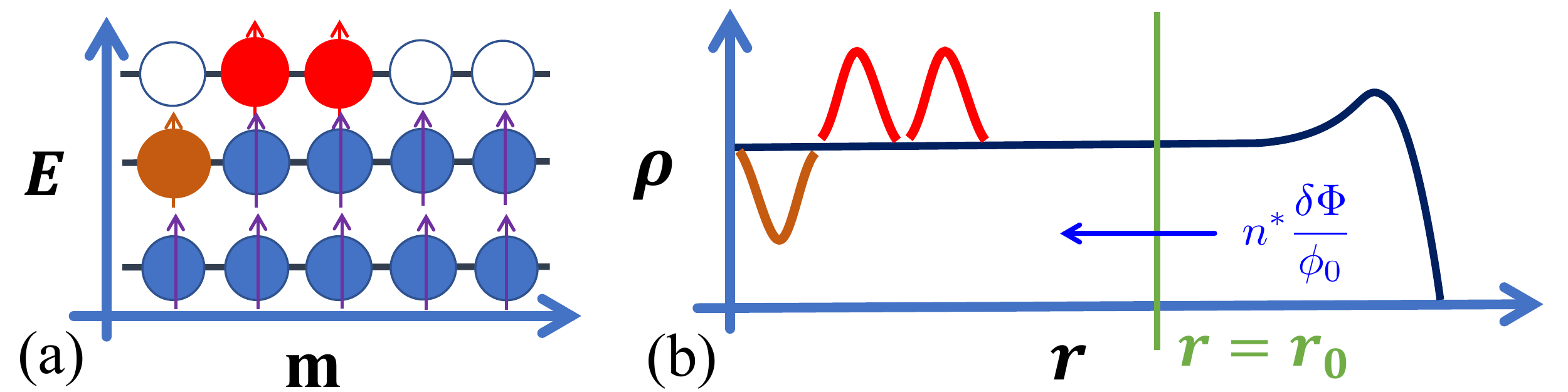}
\caption{ Schematic of the derivation of quantized charge of excitations using the self-consistency requirements. (a) Relative to the ground state with effective filling factor $n^*=2$, an anyon colored in red denotes a quasiparticle, and removal of an anyon (for example, the one colored brown) produces a quasihole in the spectrum of KS orbitals. (b) When the total flux inside the region $r<r_0$ increases by an amount $\delta \Phi$, $n^*\frac{\delta \Phi}{\phi_0}$ orbitals flow into this region.}\label{deriveQ}
\end{figure}

The ``integer" quantum Hall effect of anyons in an external magnetic field $B_{\rm ext} \vec{e}_{\rm z}$ calculated using the DFT method has been shown in the previous section. In this section, we focus on its topological properties including quasiparticle charge and statistics, as well as Hall conductance. The discussion in this section is largely independent of the results from previous sections that focus on the ground state density and energy.

The results derived below can be anticipated from the trial states that we proposed in Sec.~\ref{Trialsection}, which are closely analogous to the unprojected wave functions $\prod_{j<k}(\bar{z}_j-\bar{z}_k)^{2p}\Phi_{n^*}$ for composite fermions, where for $n^*>0$ ($n^*<0$) the wave function $\Phi_{n^*}$ represent $n^*$ filled LLs in positive (negative) magnetic field. Let us recall certain properties of these wave function~\cite{Jeon04,Jeon03b,Jain07}. (i) The factor $\prod_{j<k}(\bar{z}_j-\bar{z}_k)^{2p}$ attaches $2p$ flux quanta in the {\rm negative}-z direction. (ii) The filling factor $\nu$ is given by $\nu=n^*/(2pn^*+1)$. (iii) The magnitude of the quasiparticle charge is given by $Q=\frac{1}{2pn^*+1}$. (iv) The magnitude of the Hall conductance is given by $\sigma_{H}=\frac{n^*}{2pn^*+1}$. (v) The braid statistics of a quasihole is $\alpha^*=\frac{2p}{2pn^*+1}$.

In the present case, we are considering anyons carrying flux $\alpha \phi_0$ pointing in the positive direction, moving in an external magnetic field $B_{\rm ext}$. Their wave functions have the form 
\be
\Psi^{\rm trial}_{\nu}=[\Delta]^\alpha \Phi_{n^*}
\ee
where $\Phi_{n^*}$ must be chosen so as to produce $\Psi^{\rm trial}_{\nu}$ in the external magnetic field $B_{\rm ext}$. By analogy to the composite-fermion theory, we can make the following statements: (i) The factor $[\Delta]^\alpha=\prod_{j<k}(\bar{z}_j-\bar{z}_k)^{*\alpha}$ attaches $\alpha$ flux quanta in the {\rm positive}-z direction. (ii) The filling factor $\nu$ is given by $\nu=n^*/(1-\alpha n^*)$. (iii) The magnitude of the quasiparticle charge is given by $Q=\frac{1}{1-\alpha n^*}$. (iv) The magnitude of the Hall conductance is given by $\sigma_{H}=\frac{n^*}{1-\alpha n^*}$. (v) The braid statistics of a quasihole excitation is $\alpha^*=\frac{-\alpha}{1-\alpha n^*}$.
We see below that the results for anyons in Eqs.~\ref{ChargeQuantize}, ~\ref{Hallquantize} and ~\ref{StatisticsQuantize} are given precisely by these formulas. We stress that the charge and statistics of the excitations are not equal to the charge and statistics of the underlying anyons, indicating the non-triviality of even a system of noninteracting anyons.

We note that the IQHE of anyons has been considered by Ma and Zhang~\cite{Ma91}. They calculate the statistics and charge of excitations for Laughlin-like states that are amenable to plasma analogy. Their results are entirely consistent with ours. 

\subsection{Charge of excitations}

We now show how these quantities can be derived using the DFT method, paying attention to the self-consistency condition. The arguments below rely on adiabatic assumption, and thus are valid only in the presence of a finite gap; as a result, they do not apply to anyons in zero external magnetic field, where a massless ``phonon" mode exists~\cite{Laughlin88science,Fetter89,Chen89,Hanna89,Xie90}. In other words, the discussion below assumes that the effective filling factor $n^*$ is different from $n=1/\alpha$. We also exclude the situation of $n^*=0$, where a Fermi sea behavior is expected and the gap again vanishes.

We consider a finite region $S$ of radius $r_0$ and area $\pi r^2_0$ around the center, in which $N(r_0)$ anyons are enclosed (Here $N(r_0)$ is not necessarily an integer). We assume that in the ground state obtained by DFT, $n^*$ Landau levels are filled everywhere inside the region $S$. We denote the ground state density as $\rho_{\rm g}$. Suppose we now create $N_p$ quasiparticles and $N_h$ quasiholes deep inside $S$, where a quasiparticle is an anyon in a higher Landau level and a quasihole is a missing anyon from a lower Landau level. We ask what is the net increase of particle number, $\delta  N(r_0)$, inside $S$.

From the DFT calculation (below), this is given by
\be
    \delta N(r_0)=\int_0^{r_0} [\rho(r')-\rho_{\rm g}(r')]2\pi r'dr'\;,\label{deltaNdef}
\ee
where $\rho(r)$ is the density of the state with $N_p$ quasiparticles and $N_h$ quasiholes.
Fig.~\ref{deriveQ} (a) shows schematically a state with $N_p=2$ and $N_h=1$. Such a state can be produced in DFT through either constrained DFT (see below) or by placing disorder potential. 
We choose $S$ to be large enough that the quasiparticles and quasiholes are completely contained inside $S$. 

The quantity $\delta N$ can be obtained without a full DFT calculation as follows. We define $\delta \Phi$ as the net increase in the total flux attached to anyons inside $S$. Since each anyon carries $\alpha\phi_0$, the following relation is obvious, 
\be
    \delta N \alpha \phi_0=\delta \Phi. \label{selfOne}
\ee
In the meanwhile, $\delta \Phi$ also changes the boundary condition for KS orbitals. As a result, $n^*\frac{\delta \Phi}{\phi_0}$ orbitals will have moved across the boundary of radius $r_0$ into $S$ due to the spectral flow\cite{Laughlin81,Halperin82}. (This correction is incorporated in the DFT calculation by the  requirement of self-consistency.)  Therefore, we have
\be
    \delta N=N_p-N_h+n^*\frac{\delta \Phi}{\phi_0}.\label{selfTwo}
\ee
Combining  Eq.~\ref{selfOne} and ~\ref{selfTwo}, we obtain
\be
    \delta N=\frac{1}{1-n^*\alpha}(N_p-N_h).\label{ChargeQuantize1}
\ee
This shows that each excitation has a charge deficiency or excess of 
\be
Q=\frac{1}{1-n^*\alpha}.\label{ChargeQuantize}
\ee 
Certain examples calculated using Eq.~\ref{ChargeQuantize} are shown in Table~\ref{tab:table}.

\begin{figure}[t]
\includegraphics[width=\columnwidth]{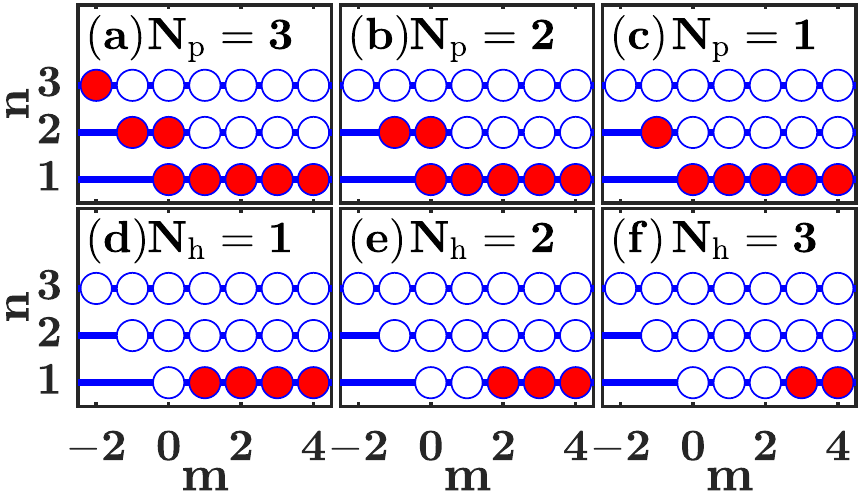}
\includegraphics[width=\columnwidth]{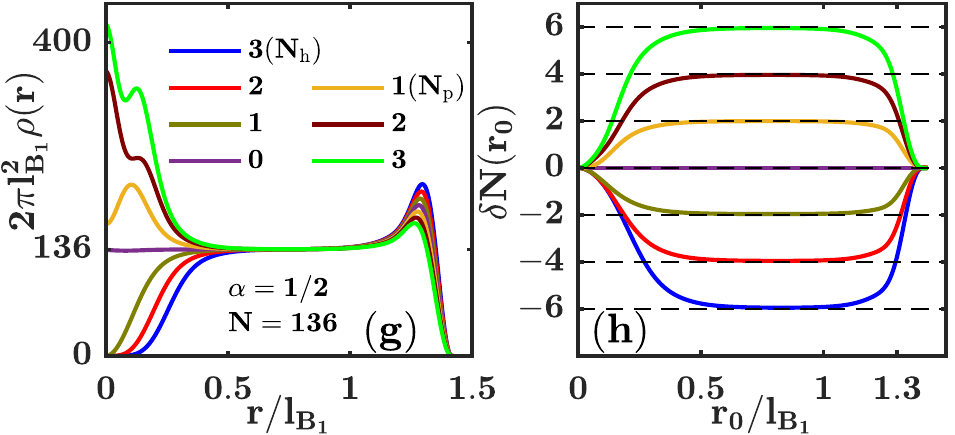}
\caption{The quantizations of charge of excitations for anyons with $\alpha=1/2$ in a hard wall potential. The excitations are created using the constrained DFT whose occupation configurations are partially shown in panels (a-f) (only the central region with excitations is shown). The occupied orbitals are marked in red and $N_{\rm p} (N_{\rm h})$ counts the number of quasiparticles (quasiholes). The quantities $n$ and $m$ are, respectively, the Landau level index and angular momentum number of the KS orbitals. (g) Density of the ground state as well as the excited states obtained from constrained DFT with occupations shown in panels (a-f). The quantity $2\pi l^2_{B_{\rm 1}}\rho(r)$, where $l_{B_{\rm 1}}$ is defined in Eq.~\ref{B1Definition}, equals $N$ in the mean field approximation. (h) This panel shows $\delta N(r_0)$, total charge accumulation inside a region of radius $r_0$, calculated using Eq.~\ref{CQFull}. It is quantized in multiples of $2$ per excitation, consistent with Eq.~\ref{ChargeQuantize}  (see the first example in Table~\ref{tab:table}). The particle number of the system is taken to be $N=136$. The external magnetic field is chosen to be $B_{\rm ext}=\alpha N B_1$, which for a uniform density state, would produce $n^*=2\alpha=1$. (In our calculation, density is not uniform.) }\label{CalcuQ}
\end{figure}

As a test of the first example in Table~\ref{tab:table}, we calculate $\delta N$ using DFT, with results shown in Fig.~\ref{CalcuQ}, where we consider a system of $N=136$ anyons with $\alpha=1/2$. We have applied an external magnetic field $B_{\rm ext}=N B_{1}/2$ such that in the ground state only $n^*=1$ Landau levels are occupied in the bulk. We construct quasiparticles and quasiholes using the constrained DFT, with the fixed occupation configurations shown in Fig.~\ref{CalcuQ}(a-f). Only the region near the center is shown in the figure; the lowest Landau is occupied until the edge of the system. (Excitations can also be generated by introducing external charged impurities\cite{Hu19}; however, a constrained DFT is convenient to control the number of quasiparticles $N_{\rm p}$ and also to achieve fast convergence at zero temperature.) The density distributions generated by the above configurations are shown in Fig.~\ref{CalcuQ}(g). The panel (h) shows the excess charge in a region of radius $r_0$. For large enough $r_0$, but for $r_0$ less than the sample size, the excess charge takes the quantized value predicted above. 

\subsection{Hall conductance}

\begin{figure}[t]
\includegraphics[width=\columnwidth]{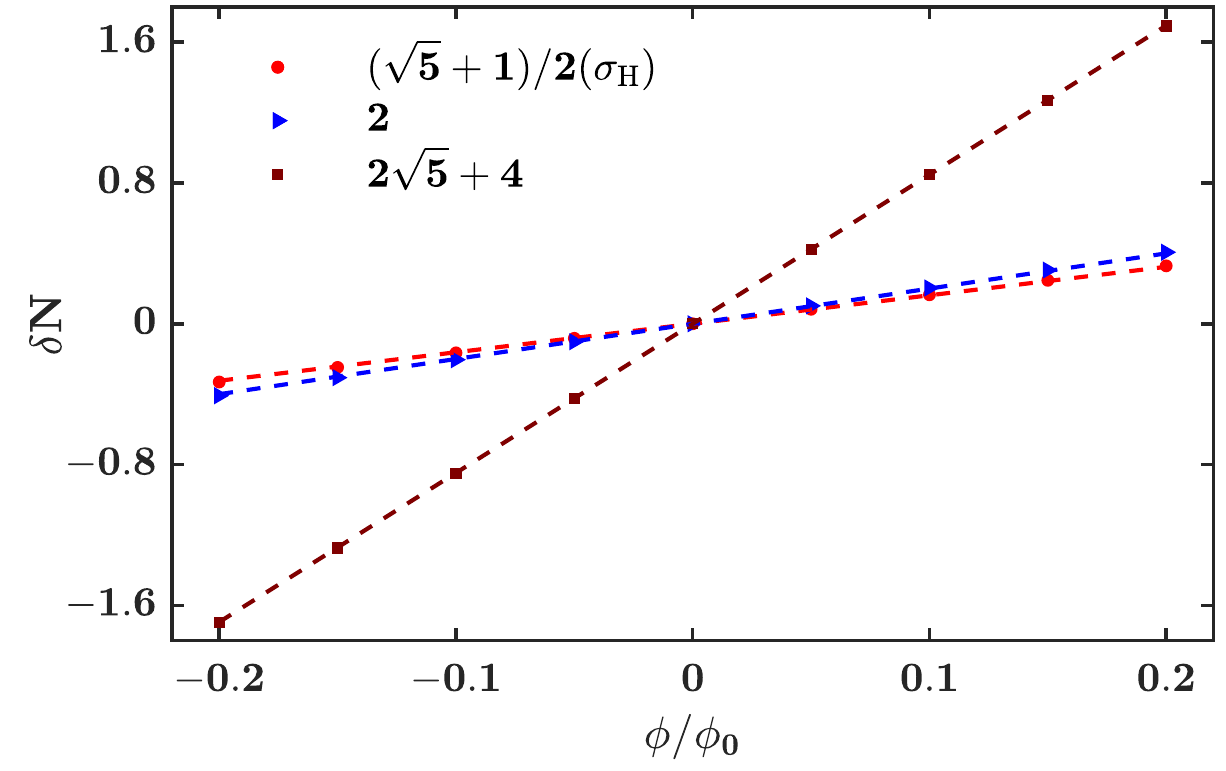}
\caption{
This figure shows the particle accumulation $\delta N$ around the origin as a function of the external flux $\phi$ through the center of the potential well. The parameters of the systems considered are given in Table~\ref{tab:table}. The slope of $\delta N$ versus $\phi$ gives $\sigma_{\rm H}$ in units of $e^2/h$. The numerical results (dots, triangles and squares) are consistent with the dashed lines which are plotted using Eq.~\ref{CQFull} by setting $N_{\rm p}=N_{\rm h}=0$ and correspond to $\sigma_{\rm H}$ shown on the figure. A constrained DFT is applied with the constrained orbitals being the ground state occupation in the absence of the external flux. $\delta N$ is determined as the plateau value from the $\delta N(r_0)$ versus $r_0$ curve, as shown in Fig.~\ref{CalcuQ}(h). The system contains $N=600$ particles.
}\label{CalcuHall}
\end{figure}

Next we show the quantization of Hall conductance, using the standard gauge argument~\cite{Laughlin81}. For this purpose, we insert adiabatically an additional external flux $\phi$ through the center, so that Eq.~\ref{selfOne} is modified into
\be
    \delta N \alpha \phi_0+\phi=\delta \Phi \label{selfThree}.
\ee
Here the value of $\phi/\phi_0$ is continuously tunable\cite{Arovas84}. The solution in Eq.~\ref{ChargeQuantize1} is modified into  
\be
    \delta N=\frac{1}{1-n^*\alpha}(N_p-N_h)+\frac{n^*}{1-n^*\alpha}{\phi \over \phi_0}\;.\label{CQFull}
\ee
In the absence of excitations, this means that for each flux quantum through the center, we will have $\delta N=\frac{n^*}{1-n^*\alpha}$ particles transfer across the boundary of $S$. We thus obtain a Hall conductance of
\be 
\sigma_{\rm H}=\frac{n^*}{1-n^*\alpha},\;\label{Hallquantize}
\ee
which is nothing but the filling factor $\nu$ at $\nu^*=n^*$, as seen from Eq.~\ref{RealFillingnu*}. That the Hall conductivity is given by the filling factor $\nu$ is expected from the observation that the Hall conductivity is determined by the collective motion of the system rather than the relative motion of particles around one another; the situation is closely analogous to the case of composite fermions, for which the Hall conductivity is determined by the real filling factor $\nu$ rather than the effective filling factor of composite fermions. In Table~\ref{tab:table} we show $\sigma_{\rm H}$ for certain values of the statistics parameter $\alpha$, including at the $\alpha=(\sqrt{5}+1)/2$ (this value is taken to stress that our discussion here is valid for arbitrary $\alpha$). In Fig.~\ref{CalcuHall}, we provide numerical tests of  the three examples in Table~\ref{tab:table} using the DFT method, where we perform a linear response measurement by calculating $\delta N$ versus a small amount of external flux $\phi$ inserted through the center of the potential well. The Hall conductance $\sigma_{\rm H}$ can be extracted from the slope of $\delta N$ versus $\phi$. 
We assume that $\phi$ is small enough that we have $N_p=N_h=0$ throughout, which allows us to 
perform a constrained DFT calculation. We note that the Kohn-Sham orbitals themselves evolve as the test flux inserted at the center alters the boundary conditions.The change of $\delta N$ is consistent with the dashed lines which are plotted using Eq.~\ref{CQFull} by setting $N_{\rm p}=N_{\rm h}=0$. We emphasize that $\delta N$ is independent of the radius $R$ as well as $N$ for a sufficiently large system.

\begin{table}[h!]
  \begin{center}
    \begin{tabular}{l|l|c|r|r} 
      \hline
      \hline
      $\alpha$ & $B_{\rm ext}/B_{\rm eff}$ & $n^*$ & $Q/e$ & $\sigma_{\rm H}/(e^{2}/h)$ \\
      \hline
      \hline
      1/2 &1  & 1 & 2 & 2\\
$(3-\sqrt{5})/2$ & $(\sqrt{5}-1)/2$ & 1 & $(\sqrt{5}+1)/2$& $(\sqrt{5}+1)/2$\\
$(3-\sqrt{5})/2$ & $(\sqrt{5}-2)/2$ & 2 & $\sqrt{5}+2$& $2\sqrt{5}+4$\\
      \hline
      \hline
    \end{tabular}
    \caption{This table lists, for convenience, various quantum numbers for certain values of the statistics parameter $\alpha$ and $B_{\rm ext}$ that we have considered. These are obtained using Eqs.~\ref{ChargeQuantize} and \ref{Hallquantize}. }  \label{tab:table}
  \end{center}
\end{table}

\begin{figure}[t]
\includegraphics[width=\columnwidth]{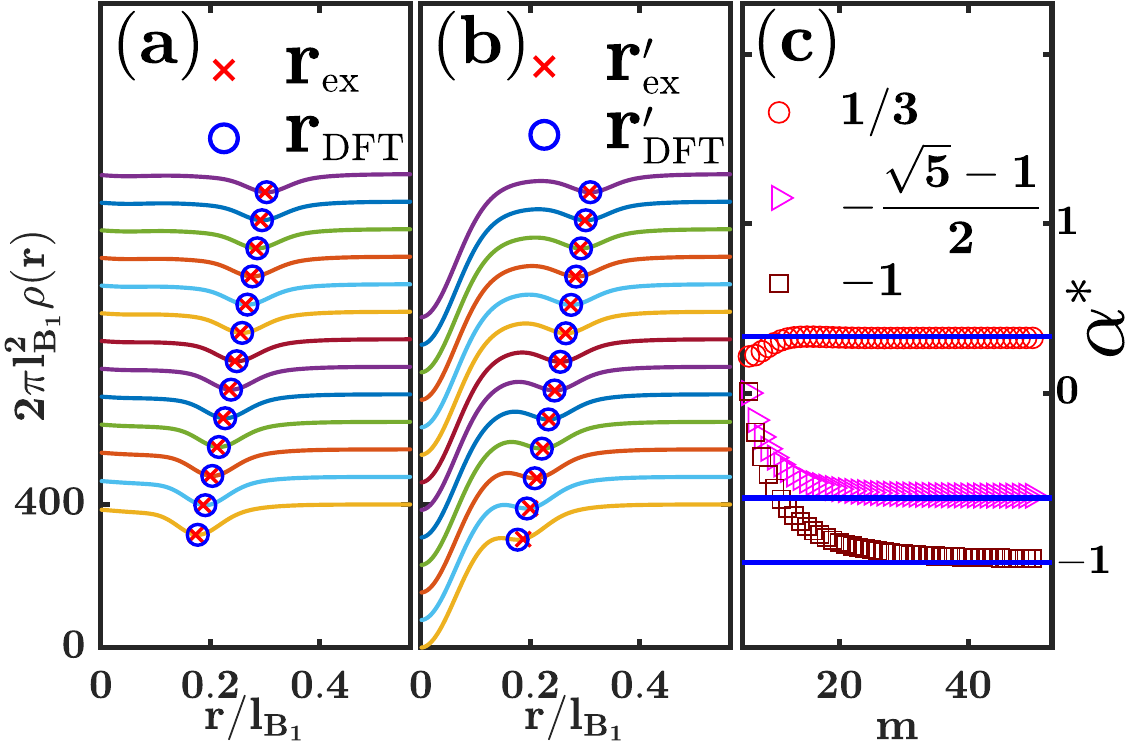}
\caption{The braid statistics between the quasiholes is obtained through the shift in the orbit of a test quasihole when another quasihole is inserted in the interior. 
(a) The density profile of a system with a single quasihole in angular momentum $m$ orbital, where $m$ ranges from $6$ to $20$ for curves from the bottom to the top, and $r$ is the distance from the center measured in units of $l_{B_1}$ (see Eq.~\ref{B1Definition} for the definitions of $l_{B_1}$ and $B_1$). Each successive curve has been shifted up vertically for clarity. We assume anyons with statistics parameter $\alpha=1/2$ with $n^*=1$. 
(b) Same as in (a) but in the presence of another quasihole with angular momentum $m=0$ at the origin.
The expected positions of the test quasihole in (a) $r_{\rm ex}=\sqrt{2m}l_{B}$ and (b) $r^\prime_{\rm ex}=\sqrt{2(m-\alpha^*)}l_{B}$ are indicated by a red cross, where $l_{B}=\sqrt{\frac{\hbar c}{e|B|}}$ and $B=n^*  N B_1$ is the mean field value of magnetic field in the system. The corresponding positions in the DFT calculation are obtained from the locations of the local minima near the expected positions; these are indicated by a blue circle. 
(c) The statistics parameter $\alpha^*\equiv (r^2_{\rm DFT}-r^{\prime 2}_{\rm DFT})/2l^2_{B}$ is calculated from the DFT data for each angular momentum $m$ ranging from 6 to 50, shown by brown squares. It approaches the expected value of $\alpha^*=-1$ for sufficiently large $m$. In this panel, we also show the statistics parameter calculated for anyons with $\alpha=(3-\sqrt{5})/2$ and $n^*=1$  (magenta triangles), which approaches the expected statistics $\alpha^*=-(\sqrt{5}-1)/2$; and also for anyons with $\alpha=-1/2$ at $n^*=1$ (red circles), which approaches the expected value $\alpha^*=1/3$. (The expected values are marked by the horizontal blue lines.)
All calculations are done on a system with $N=400$ particles.}\label{StatisticsFig}
\end{figure}

\subsection{Statistics of excitations}

We next consider the braiding statistics of the excitations by DFT, closely following the treatment in Ref.~\onlinecite{Hu19}. According to Eq.~\ref{ChargeQuantize}, when a quasihole is created, it introduces into the system an extra amount of flux
\be
\alpha^*=-\alpha Q = -\frac{\alpha}{1-n^*\alpha}\;, \label{StatisticsQuantize}
\ee
in units of $\phi_0$, where the minus sign is due to the fact that we take flux along the $\vec{e}_{\rm z}$ direction to be positive. Other particles that orbit around this quasihole will feel the missing flux, and $\alpha^*$ can be interpreted as the braiding statistics of the quasihole when taking the ground state of anyons as the reference vacuum. One consequence of the braiding statistics is that the radius of a quasihole orbital depends on how many other quasiholes it encloses. The quasihole statistics is thus related to the shift of its orbital when another quasihole is inserted in the interior. In the absence of any other quasiholes, the expected position of a test quasihole in angular momentum $m$ orbital is $r_{\rm ex}=\sqrt{2m}l_{B}$, where $l_{B}=\sqrt{\frac{\hbar c}{e|B|}}$ and $B=n^*  N B_1$ is the mean field value of magnetic field. When another quasihole is  inserted at the center, the position of the test quasihole is expected to shift to $r^\prime_{\rm ex}=\sqrt{2(m-\alpha^*)}l_{B}$. Let us now look at our constrained DFT calculations. 
In Fig.~\ref{StatisticsFig} we show the position of a test quasihole with angular momentum $m$.  The positions $r_{\rm DFT}$ and $r^\prime_{\rm DFT}$ of the test quasihole are determined by the local density minimum near the expected position. The numerical positions are seen to be consistent with the expected values when the two quasiholes are far apart. We also extract the statistics parameters using $\alpha^*\equiv (r^2_{\rm DFT}-r^{\prime 2}_{\rm DFT})/2l^2_{B}$. It is seen to be quantized at the expected values obtained from Eq.~\ref{StatisticsQuantize} when $m$ is large enough.

The quantized charge and statistics of excitations and the Hall conductance for non-interacting anyons have been derived within our model that neglects exchange correlation (xc) effects beyond those included through the effective magnetic field. One can ask if these topological properties survive when the exact xc potential (which is not known) is taken into consideration. We expect that a more accurate xc potential will influence the detailed density profile locally and possibly change the radius of the excitations, but, so long as the influence of xc potential is local, it will not affect the topological properties of the state. This is because the above proofs do not depend on the details of the ground state density deep inside the region $S$, but only require a uniform occupation number of $n^*$ in the ground state. The same argument also implies that the topological properties are robust against weak disorders that do not close the gap.

Our above analysis can be generalized straightforwardly to fractional quantum Hall effect of anyons~\cite{Canright89,Canright89a,Lee89,Lee91,Ma91,Ciftja05} by analogy to the CF theory, i.e. by attaching additional $2p$ flux quanta to each anyon. What interactions will produce such states 
is beyond the scope of the paper, where we only consider non-interacting anyons. 

\begin{figure}[t]
\includegraphics[width=\columnwidth]{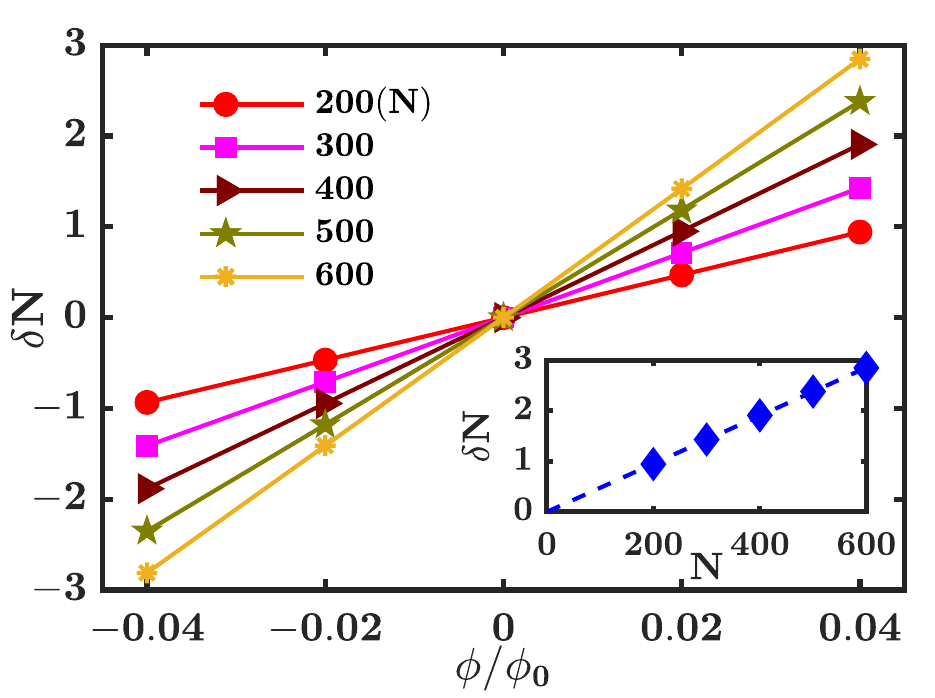}
\caption{The particle accumulation $\delta N$ around the origin in response to an external flux $\phi$ for anyons with $\alpha=1/2$ in the absence of an external magnetic field. Results for systems with different particle number $N$ are compared. The large slopes indicate that $\delta N$ changes rapidly to a small $\phi$, and the slope is proportional to the particle number $N$. The inset shows $\delta N$ versus particle number $N$ for a fixed $\phi=0.04\phi_0$ and the dashed line is a linear fitting. The effective filling factor is $n^*=2$ and $\delta N$ is calculated using Eq.~\ref{deltaNdef} in a constrained DFT.} \label{CalcuSigmaB0}
\end{figure}

It is interesting to ask how the considerations in this section relate to the possible superconducting behavior at zero external magnetic field. Note that the charge in Eq.~\ref{ChargeQuantize} diverges when we take $\alpha=1/n$ and $n^*=n$, which corresponds to $B_{\rm ext}=0$. As noted earlier, there is no real gap for these parameters, and hence, strictly speaking, our considerations above do not apply. One may nonetheless ask what a self-consistent solution of the KS equation gives. 
Within our constrained DFT, we have obtained the solution of the DFT equations in the presence of a point flux $\phi$ at the origin in Fig.~\ref{CalcuSigmaB0}, which shows the particle accumulation $\delta N$ versus $\phi$ for different particle numbers $N$. (We note that in this situation, the singular test flux $\phi$ effectively modifies Eq.~\ref{PolarHw} by $\alpha N(\bar{r})\rightarrow \alpha N(\bar{r})+\phi/\phi_0$, thus leaving the KS solutions independent of the radius $R$ of the hard wall potential.) A large slope of $\delta N$ versus $\phi$ is observed for small values of $\phi$. This means that the system is rather sensitive to perturbation due to an external flux when $B_{\rm ext}=0$. From the density profile, it is found that its charge distribution is rather delocalized. More interestingly, the slope of $\delta N$ versus $\phi$ is proportional to $N$, as shown in the inset of Fig.~\ref{CalcuSigmaB0}. 
This is reminiscent of the extensive response to an external flux in a superconductor\cite{Byers61,Canright89,Canright89a}. It is to be contrasted with the results in Fig.~\ref{CalcuHall} for the quantum Hall states of anyons, where the slope is independent of the particle number. 

When a quasiparticle or a quasihole is added to the system, the ``internal" flux attached to anyons causes a nonperturbative reorganization of the solutions even without any additional external flux. However, the solution for the KS orbitals does not change when a quasiparticle is inserted along with an equal amount of external flux $\phi=- \phi_0/n$ to perfectly screen the flux attached to the quasiparticle. This follows because insertion in this fashion does not change the effective magnetic field or the boundary conditions for any KS orbitals. As the external flux is varied away from this special value, $\delta N$ again changes very rapidly.  The relation between an excitation and a flux $\phi_0/n$ is also a signature of the equivalence between the charged particles and vortices in anyon superconductivity~\cite{Chen89}.

\section{Conclusions}
\label{sec:Conclusions}

We have investigated the system of noninteracting anyons through the Kohn-Sham density functional theory, which takes into account of the attached gauge flux in a self-consistent fashion. This work follows a previous work on composite fermions~\cite{Hu19}. 

We compare the energy obtained from our DFT calculation with exact energies known for small systems and find qualitative and semi-quantitative agreement, especially in the vicinity of the fermionic statistics. We also show that our DFT model is consistent with a class of trial wave functions that are derived by an analogy with the composite fermion theory. Our method provides an understanding of the existing exact results and also a self-consistent way to calculate the ground state properties of the many-anyon system in the thermodynamic limit. In particular, we find the Meissner-like effect, which is a signature of anyon superconductivity, recovered in the thermodynamic limit. We find a nonperturbative response to an external flux for anyons in zero external magnetic field, which is also an expected behavior from a superconductor. We also determine the quantizations of the charge and statistics of the excitations and of the Hall conductance using the self-consistent condition within our DFT model.  

Our self-consistent method goes beyond the mean field treatment of anyons. It can in principle be further improved by including the exchange-correlation interaction, which has been neglected in our work. We expect that many topological quantities are not sensitive to the exchange-correlation interaction so long as its effect is weak and local.

\textit{Acknowledgement}:
We thank G. J. Sreejith and Yinghai Wu for helpful discussions. The work at Penn State was made possible by financial support from the US Department of Energy under Grant No. DE-SC0005042. Y. H. acknowledges partial financial support from China Scholarship Council. The numerical calculations were performed using Advanced CyberInfrastructure computational resources provided by The Institute for CyberScience at The Pennsylvania State University. We thank the the Indian Institute Science, Bangalore, where part of this work was performed, for their hospitality, and the Infosys Foundation for making the visit possible. We also thank International Centre for Theoretical Sciences, Bangalore, for its hospitality and support during the workshop `Novel phases of Quantum Matter', where this work was initiated.

\appendix

\section{Magnetic-field DFT for fermions with gauge-interaction}\label{GeneralHK}

We rewrite Eq.~\ref{FullHF} as
\ba
&\mathcal H_F=\sum_i\left[\frac{1}{2M}\left(\boldsymbol{\pi}_i+\frac{e}{c}\vec{a}_i\right)^2+V_{\rm ext}(\vec{r}_i)\right]\;,\label{FullHFrewrite1}
\ea
where we have defined the mechanical momentum $\boldsymbol{\pi}_i=\vec{p}_i-\frac{q}{c}\vec{A}_{\rm ext}(\vec{r}_i)$.  To see explicitly the interaction we rewrite the kinetic energy as\cite{Xie90,Chou91}:
\begin{eqnarray}
     &\sum_i\frac{1}{2M}\left(\boldsymbol{\pi}_i+\frac{e}{c}\vec{a}_i\right)^2=T_{1}+T_{2}+T_{3}\;,\\
& T_{1}=\sum_i\frac{1}{2M}\boldsymbol{\pi}_i^2,\\
& T_{2}=\sum_{i, j}\frac{1}{2M}\left[2\frac{e}{c}\vec{a}_{ij}\cdot\boldsymbol{\pi}_i+(\frac{e}{c})^2\vec{a}^2_{ij}\right],\\
& T_{3}=\sum_{i, j, k}\left[ \frac{1}{2M}(\frac{e}{c})^2\vec{a}_{ij}\cdot\vec{a}_{jk}\right],\\
&\vec{a}_{ij}=\frac{\alpha\phi_0}{2\pi}\left(\frac{y_i-y_j}{r^2_{ij}},\frac{x_j-x_i}{r^2_{ij}}\right)(1-\delta_{ij})\;,   \label{interactingKinetic}
\end{eqnarray}
where $T_1$ is the kinetic energy in a magnetic field and $T_2$ and $T_3$ are two-body and three-body interaction terms. 
Despite the formal complexity of the interaction, the standard proof of Hohenberg-Kohn theorems in magnetic-field DFT goes through in essentially the same way as in the familiar situation of Coulomb interaction~\cite{Giuliani08,Iyetomi89,Iyetomi89a,Gezerlis10}. The first theorem states that the 
 external vector potential $\vec{A}_{\rm ext}(\vec{r})$ and the ground state density $\rho_{\rm GS}(\vec{r})$ of $\mathcal H_F$ in Eq.~\ref{FullHF} uniquely determine the external scalar potential $V_{\rm ext}(\vec{r})=V_{\rm ext}[\vec{A}_{\rm ext},\rho_{\rm GS}]$. The second theorem states that the total energy of the system is a functional of the density. That implies the existence of $F^{\alpha}[\rho, \vec{A}_{\rm ext}]$ defined in Eq.~\ref{KSstartotalEK}.

We note here in passing that for external potentials of the form $V_{\rm ext}=M\omega^2r^2/2$, the Hamiltonian in Eq.~\ref{FullHFrewrite1} can be written as $\mathcal H_F=H_{c}+H_{rel}$, where  $H_{c}$ corresponds to the center of mass (CoM) motion and $H_{\rm rel}$ to the internal relative motion. Here,
\be
H_{c}=\frac{1}{2MN}\boldsymbol{\pi}^2_c+\frac{1}{2}MN\omega^2\vec{r}^2_c \label{separationCoM}
\ee
and 
\be
H_{rel}=\frac{1}{2MN}\sum_{i<j}(\boldsymbol{\pi}_i-\boldsymbol{\pi}_j)^2_c+\frac{M\omega^2}{2N}\sum_{i<j}(\vec{r}_i-\vec{r}_j)^2+T_2+T_3\;,
\ee
where $\vec{r}_c=(\sum_i\vec{r}_i)/N$ is the CoM and $\boldsymbol{\pi}_c=\sum_i \boldsymbol{\pi}_i$ is the CoM momentum. The terms $T_2$ and $T_3$ involve only relative motion:
\begin{eqnarray}
& T_{2}=\sum_{i<j}\alpha\hbar(L_{ij}+\alpha\hbar)/|\vec{r}_i-\vec{r}_j|^2,\\
& T_{3}=\sum_{i}\sum_{ j< k,j\neq i,k\neq i}\alpha^2\hbar^2\frac{\vec{r}_i-\vec{r}_j}{|\vec{r}_i-\vec{r}_j|^2}\cdot\frac{\vec{r}_i-\vec{r}_k}{|\vec{r}_i-\vec{r}_k|^2},
\end{eqnarray} 
where $L_{ij}$ is the relative angular momentum:
\be
L_{ij}=(\vec{r}_i-\vec{r}_j)\times (\boldsymbol{\pi}_i-\boldsymbol{\pi}_j)\;.
\ee

\section{Modified trial wave functions}\label{regularize}

In Refs.~\onlinecite{Lundholm17,Lundholm13}, a ``regulator'' to the anyon trial wave functions has been proposed to obtain better variational states. The idea is to further multiply the anyon trial wave function by a regularizing symmetric function $S$:
\be
S(d,\gamma)=\prod_{i<j}\frac{|z_{ij}|^{2\gamma}}{(d^2+|z_{ij}|^2)^{\gamma}}\;,\label{SymRegulator}
\ee
and search for an energy minimum in the parameter space of $\gamma$ and $d$.

\begin{figure}[t]
\includegraphics[width=\columnwidth]{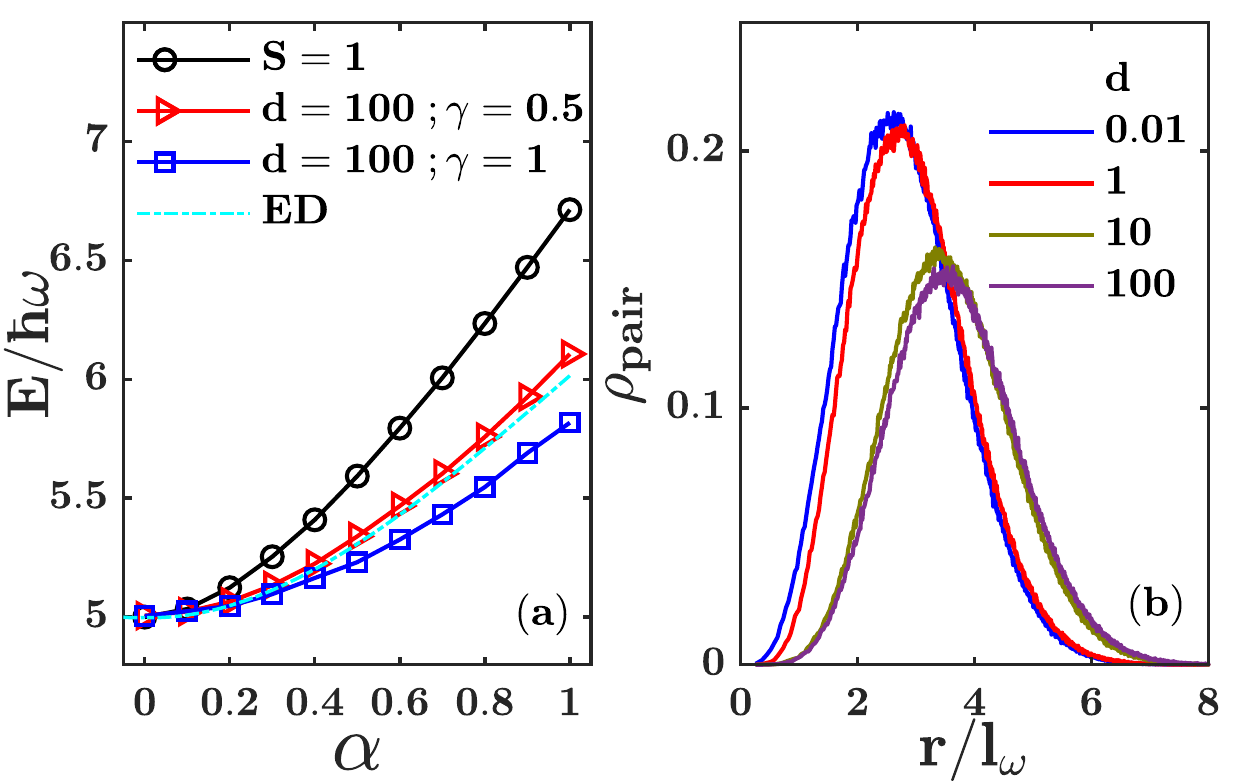}
\caption{(a) Energy of the trial wave function $S(d,\gamma)[\Delta_N]^\alpha\Phi_E$ in a parabolic potential $V_{\rm ext}=M\omega^2r^2/2$ versus the statistical parameter $\alpha$. $\Phi_E$ is the wave function of fermions in the configuration shown in panel $E$ of Fig.~\ref{ConstrainOccu}. The regulator $S(d,\gamma)$ is defined in Eq.~\ref{SymRegulator}. The black, red and blue curves show the energy of the trial wave function for a few choices of parameters shown on the figure. The cyan curve shows the energy obtained from exact diagonalization (ED); we stress that this curve represents the ground state for small $\alpha$, but an excited state in general. (b) Evolution of the pair correlation $\rho_{pair}$ of the trial wave function $S(d,\gamma)[\Delta_N]^\alpha\Phi_E$ versus $d$ with $\gamma=0.5$. The quantity $\rho_{pair}$ is plotted in units of $l^{-2}_{\omega}$.}\label{TrialTune}
\end{figure}

As an example, we consider the wave function corresponding to panel $E$ in Fig.~\ref{ConstrainOccu}. Previously we used the trial wave function $[\Delta_N]^\alpha\Phi_E$ for this state. Fig.~\ref{TrialTune}(a) reproduces the energy of this wave function (black curve), along with the energy of the exact eigenstate (cyan curve). We stress that the cyan curve is the ground state for $\alpha=0$ (the fermionic point) and for small values of $\alpha$, but for larger $\alpha$ it is an excited state. We also show the energy of $S(d,\gamma)[\Delta_N]^\alpha\Phi_E$ for certain values of $d$ and $\gamma$. It is evident that the energy can be reduced by multiplication by $S(d,\gamma)$. (One need not be disturbed by the fact that the energy goes below the cyan curve, as that curve is not the ground state except for small $\alpha$.) The effect of $d$ is to tune the short-distance (on the order of $ d$ or smaller) correlations between particles, which can be seen in the pair distribution of anyons in Fig.~\ref{TrialTune}(b). Since $S$ does not produce additional phases, 
and also does not influence the long-distance inter-particle correlations, 
we expect that it will also not influence the topological properties of the anyons. 

We mention here other studies of this system.
Ref.~\onlinecite{Chin92} has given 
a different class of trial wave functions for three anyons, which are in good agreement with exact results. We have compared our trial wave functions with theirs. For $\alpha=\gamma=0.5$, the overlap of our state with that in Ref.~\onlinecite{Chin92} decreases as we increase $d$ from 0 to 1000, even as their energies come closer.  Ref.~\onlinecite{Chitra92} uses a semiclassical approximation for the cyan curve discussed above.

\section{Semiclassical approximation for fermions density in a parabolic potential}\label{classicalDensity}

\begin{figure}[t]
\includegraphics[width=\columnwidth]{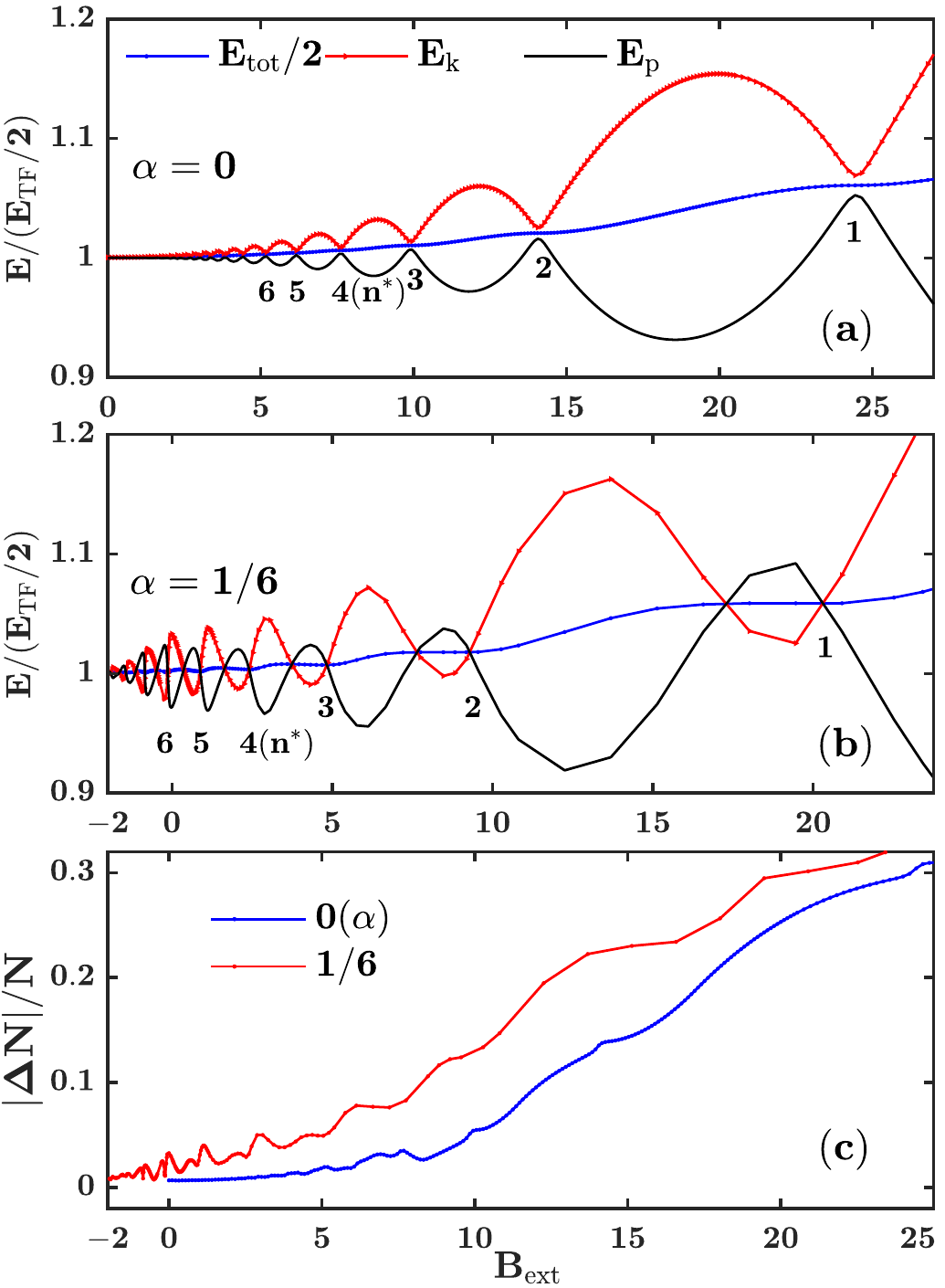}
\caption{ Comparison of DFT energies for fermions (panel a) and anyons with $\alpha=1/6$ (panel b) with those obtained from the semiclassical Thomas-Fermi approximation as a function of $B_{\rm ext}$. Here $E_{\rm tot}$ is the total energy, $E_{\rm k}$ is the kinetic energy, $E_{\rm p}$ is the potential energy, and $E_{\rm TF}$ is the total Thomas Fermi energy. $E_{\rm tot}$, $E_{\rm k}$ and $E_{\rm p}$ are calculated using DFT. The numbers $n^*$ show the effective filling factor in the bulk.  We see that $E_{\rm tot}/E_{\rm TF}$ is nearly unity for small $B_{\rm ext}$. Panel c shows $|\Delta N|$, defined in Eq.~\ref{deviationDensity}, which is the absolute deviation of the density from the Thomas Fermi value. The calculations are for a system of $N=600$ particles at temperature $k_{\rm B}\tau=0.005\hbar\omega$.}\label{SemiCla}
\end{figure}

Following Ref.~\onlinecite{Butts97}, for a large particle number $N$, we obtain the density distribution of fermions in a two-dimensional harmonic potential using the semiclassical Thomas-Fermi approximation. At zero temperature, we define a ``local" Fermi wave vector $k_{\rm F}(\vec{r})$ by
\be
\frac{\hbar^2k^2_{\rm F}(\vec{r})}{2M}+\frac{1}{2}M\omega^2r^2=E_{\rm F}, \label{ClassicalTF}
\ee  
where $E_{\rm F}$ is the Fermi energy. The density $\rho_{\rm TF}(\vec{r})$ is the area of the local Fermi sea in k-space multiplied by the density of states $(2\pi)^{-2}$,  
\be 
\rho_{\rm TF}(\vec{r})=\frac{M}{2\pi\hbar^2}(E_{\rm F}-\frac{1}{2}M\omega^2r^2),
\ee 
and $\rho_{\rm TF}(\vec{r})=0$ when $r>R_{\rm F}=(2E_{\rm F}/{M\omega^2})^{1/2}$. By requiring $\int\rho_{\rm TF}(\vec{r})d^2\vec{r}=N$, we obtain $E_{\rm F}=\hbar\omega\sqrt{2N}$ and $R_{\rm F}=2(2N)^{1/4}l_{\omega}$. To further simplify, we express $\rho_{\rm TF}(\vec{r})$ in a dimensionless form as $\bar{\rho}_{\rm TF}(\vec{r})=2\pi l^2_{\omega}\rho(\vec{r})$,
\be
\bar{\rho}_{\rm TF}(\bar{r})=\frac{\sqrt{2N}}{2}-\frac{1}{8}\bar{r}^2,
\ee
where $\bar{r}=r/l_{\omega}$.
The ground state is nondegenerate for a ``magic'' particle number $N=(m_{\rm max}+1)(m_{\rm max}+2)/2$, where $m_{\rm max}$ is the maximum angular momentum of the single particle orbitals below Fermi surface. We have $\bar{\rho}(r=0)\approx(2m_{\rm max}+3\pm 1)/4$, which is exactly the number of occupied orbitals with angular momentum $m=0$ when $m_{\rm max}$ is even/odd. By direct integrations, we obtain the total kinetic energy and total potential energy which are both half the total energy, $E_{\rm k}=E_{\rm p}=E_{\rm TF}/2$, where $E_{\rm TF}=\hbar\omega(2N)^{3/2}/3$ is the Thomas-Fermi approximation of total energy, to be differentiated from the exact total energy $\hbar\omega N\sqrt{8N+1}/3$ for a ``magic'' $N$.

The above Thomas-Fermi approximation generalizes to anyons through a replacement of Eq.~\ref{ClassicalTF} by Eq.~\ref{AnyonClassical}. However, in a crude approximation, the effect of the vector potential is irrelevant to the density and energy, for reasons mentioned in the main text. We thus expect the energy and density of anyons to be the same as those for fermions.

Of course, one may ask how accurate the Thomas-Fermi approximation is in the presence of a magnetic field. More sophisticated treatments of the Thomas-Fermi approximation can be found in Ref.~\onlinecite{Fushiki92} for homogeneous Fermi gas in the presence of an external magnetic field, and in Ref.~\onlinecite{Chitra92} for anyons treated as fermions in an effective magnetic field. Both studies show that the ground state energy is insensitive to the external or effective magnetic field. Using the DFT method, we consider the combined effects of both the effective and external magnetic field. Our results show significant deviations from the semiclassical Thomas-Fermi approximation only when the magnetic field is strong enough that particles occupy only the lowest few Landau levels. In Fig.~\ref{SemiCla}, we show how the energy and density deviates from the Thomas-Fermi approximation versus the strength of $B_{\rm ext}$.  In Fig.~\ref{SemiCla}(a), we notice that the total energy of the Thomas-Fermi approximation is valid in a large range of $B_{\rm ext}$ even when Landau levels form in the system, though the kinetic and potential energies fluctuate more dramatically. The same observation holds true for anyons [see Fig.~\ref{SemiCla}(b)]. In particular, the Thomas Fermi approximation is good in the range of $-1\leq\alpha\leq1$ and when the externally applied magnetic field is of the same order of the effective magnetic field.
We describe the deviation of density distribution from the Thomas-Fermi approximation using the following quantity: 
\be
    \Delta N=\int_0^{\infty} |\rho(r')-\rho_{\rm TF}(r')|2\pi r'dr'\;.\label{deviationDensity}
\ee
It is shown in Fig.~\ref{SemiCla}(c) that the density distribution of both fermions and anyons stick around $\rho_{\rm TF}$ for a large range of external magnetic field until only the lowest few Landau levels are occupied. We also point out that in the DFT results, both bosons ($\alpha=1$) and semions ($\alpha=0.5$) also have density distributions around $\rho_{\rm TF}$ in the absence of an external magnetic field, even though only one or two Landau levels are occupied.

%\bibliography{../../Latex-Revtex-Master/biblio_fqhe.bib}

\begin{thebibliography}{85}
\expandafter\ifx\csname natexlab\endcsname\relax\def\natexlab#1{#1}\fi
\expandafter\ifx\csname bibnamefont\endcsname\relax
  \def\bibnamefont#1{#1}\fi
\expandafter\ifx\csname bibfnamefont\endcsname\relax
  \def\bibfnamefont#1{#1}\fi
\expandafter\ifx\csname citenamefont\endcsname\relax
  \def\citenamefont#1{#1}\fi
\expandafter\ifx\csname url\endcsname\relax
  \def\url#1{\texttt{#1}}\fi
\expandafter\ifx\csname urlprefix\endcsname\relax\def\urlprefix{URL }\fi
\providecommand{\bibinfo}[2]{#2}
\providecommand{\eprint}[2][]{\url{#2}}

\bibitem[{\citenamefont{Leinaas and Myrheim}(1977)}]{Leinaas77}
\bibinfo{author}{\bibfnamefont{J.}~\bibnamefont{Leinaas}} \bibnamefont{and}
  \bibinfo{author}{\bibfnamefont{J.}~\bibnamefont{Myrheim}},
  \bibinfo{journal}{Il Nuovo Cimento B Series 11}
  \textbf{\bibinfo{volume}{37}}, \bibinfo{pages}{1} (\bibinfo{year}{1977}),
  ISSN \bibinfo{issn}{0369-3554},
  \urlprefix\url{http://dx.doi.org/10.1007/BF02727953}.

\bibitem[{\citenamefont{Wilczek}(1982)}]{Wilczek82}
\bibinfo{author}{\bibfnamefont{F.}~\bibnamefont{Wilczek}},
  \bibinfo{journal}{Phys. Rev. Lett.} \textbf{\bibinfo{volume}{49}},
  \bibinfo{pages}{957} (\bibinfo{year}{1982}),
  \urlprefix\url{http://link.aps.org/doi/10.1103/PhysRevLett.49.957}.

\bibitem[{\citenamefont{Wilczek}(1990)}]{Wilczek90}
\bibinfo{author}{\bibfnamefont{F.}~\bibnamefont{Wilczek}},
  \emph{\bibinfo{title}{Fractional Statistics and Anyon Superconductivity}}
  (\bibinfo{publisher}{World Scientific}, \bibinfo{year}{1990}), ISBN
  \bibinfo{isbn}{9789810200480},
  \urlprefix\url{http://books.google.com/books?id=vFyoQgAACAAJ}.

\bibitem[{\citenamefont{Iengo and Lechne}(1992)}]{Roberto92}
\bibinfo{author}{\bibfnamefont{R.}~\bibnamefont{Iengo}} \bibnamefont{and}
  \bibinfo{author}{\bibfnamefont{K.}~\bibnamefont{Lechne}},
  \bibinfo{journal}{Physics Reports} \textbf{\bibinfo{volume}{213}},
  \bibinfo{pages}{179} (\bibinfo{year}{1992}).

\bibitem[{\citenamefont{Lerda}(1992)}]{Lerda92}
\bibinfo{author}{\bibfnamefont{A.}~\bibnamefont{Lerda}},
  \emph{\bibinfo{title}{Anyons: Quantum Mechanics of Particles with Fractional
  Statistics}} (\bibinfo{publisher}{Springer Science \& Business Media},
  \bibinfo{year}{1992}).

\bibitem[{\citenamefont{Rao}(1997)}]{Rao92}
\bibinfo{author}{\bibfnamefont{S.}~\bibnamefont{Rao}}, \emph{\bibinfo{title}{An
  anyon primer}} (\bibinfo{publisher}{Narosa publications},
  \bibinfo{year}{1997}), ISBN \bibinfo{isbn}{978-8173191244}.

\bibitem[{\citenamefont{Khare}(2005)}]{Khare05}
\bibinfo{author}{\bibfnamefont{A.}~\bibnamefont{Khare}},
  \emph{\bibinfo{title}{Fractional statistics and quantum theory}}
  (\bibinfo{publisher}{World Scientific}, \bibinfo{year}{2005}).

\bibitem[{\citenamefont{Rao}(2017)}]{Rao17}
\bibinfo{author}{\bibfnamefont{S.}~\bibnamefont{Rao}},
  \emph{\bibinfo{title}{Introduction to abelian and non-abelian anyons}}
  (\bibinfo{publisher}{Springer Singapore}, \bibinfo{address}{Singapore},
  \bibinfo{year}{2017}), pp. \bibinfo{pages}{399--437}, ISBN
  \bibinfo{isbn}{978-981-10-6841-6},
  \urlprefix\url{https://doi.org/10.1007/978-981-10-6841-6_16}.

\bibitem[{\citenamefont{Chen et~al.}(1989)\citenamefont{Chen, Wilczek, Witten,
  and Halperin}}]{Chen89}
\bibinfo{author}{\bibfnamefont{Y.-H.} \bibnamefont{Chen}},
  \bibinfo{author}{\bibfnamefont{F.}~\bibnamefont{Wilczek}},
  \bibinfo{author}{\bibfnamefont{E.}~\bibnamefont{Witten}}, \bibnamefont{and}
  \bibinfo{author}{\bibfnamefont{B.~I.} \bibnamefont{Halperin}},
  \bibinfo{journal}{International Journal of Modern Physics B}
  \textbf{\bibinfo{volume}{3}}, \bibinfo{pages}{1001} (\bibinfo{year}{1989}).

\bibitem[{\citenamefont{Iengo and Lechner}(1991)}]{Roberto91}
\bibinfo{author}{\bibfnamefont{R.}~\bibnamefont{Iengo}} \bibnamefont{and}
  \bibinfo{author}{\bibfnamefont{K.}~\bibnamefont{Lechner}},
  \bibinfo{journal}{Nuclear Physics B} \textbf{\bibinfo{volume}{364}},
  \bibinfo{pages}{551} (\bibinfo{year}{1991}).

\bibitem[{\citenamefont{Sporre et~al.}(1991)\citenamefont{Sporre, Verbaarschot,
  and Zahed}}]{Sporre91}
\bibinfo{author}{\bibfnamefont{M.}~\bibnamefont{Sporre}},
  \bibinfo{author}{\bibfnamefont{J.}~\bibnamefont{Verbaarschot}},
  \bibnamefont{and} \bibinfo{author}{\bibfnamefont{I.}~\bibnamefont{Zahed}},
  \bibinfo{journal}{Physical review letters} \textbf{\bibinfo{volume}{67}},
  \bibinfo{pages}{1813} (\bibinfo{year}{1991}).

\bibitem[{\citenamefont{Murthy et~al.}(1991)\citenamefont{Murthy, Law, Brack,
  and Bhaduri}}]{Murthy91}
\bibinfo{author}{\bibfnamefont{M.}~\bibnamefont{Murthy}},
  \bibinfo{author}{\bibfnamefont{J.}~\bibnamefont{Law}},
  \bibinfo{author}{\bibfnamefont{M.}~\bibnamefont{Brack}}, \bibnamefont{and}
  \bibinfo{author}{\bibfnamefont{R.}~\bibnamefont{Bhaduri}},
  \bibinfo{journal}{Physical review letters} \textbf{\bibinfo{volume}{67}},
  \bibinfo{pages}{1817} (\bibinfo{year}{1991}).

\bibitem[{\citenamefont{Sporre et~al.}(1992)\citenamefont{Sporre, Verbaarschot,
  and Zahed}}]{Sporre92}
\bibinfo{author}{\bibfnamefont{M.}~\bibnamefont{Sporre}},
  \bibinfo{author}{\bibfnamefont{J.}~\bibnamefont{Verbaarschot}},
  \bibnamefont{and} \bibinfo{author}{\bibfnamefont{I.}~\bibnamefont{Zahed}},
  \bibinfo{journal}{Physical Review B} \textbf{\bibinfo{volume}{46}},
  \bibinfo{pages}{5738} (\bibinfo{year}{1992}).

\bibitem[{\citenamefont{Sporre et~al.}(1993)\citenamefont{Sporre, Verbaarschot,
  and Zahed}}]{Sporre93}
\bibinfo{author}{\bibfnamefont{M.}~\bibnamefont{Sporre}},
  \bibinfo{author}{\bibfnamefont{J.}~\bibnamefont{Verbaarschot}},
  \bibnamefont{and} \bibinfo{author}{\bibfnamefont{I.}~\bibnamefont{Zahed}},
  \bibinfo{journal}{Nuclear Physics B} \textbf{\bibinfo{volume}{389}},
  \bibinfo{pages}{645} (\bibinfo{year}{1993}).

\bibitem[{\citenamefont{Canright and Girvin}(1989)}]{Canright89}
\bibinfo{author}{\bibfnamefont{G.}~\bibnamefont{Canright}} \bibnamefont{and}
  \bibinfo{author}{\bibfnamefont{S.}~\bibnamefont{Girvin}},
  \bibinfo{journal}{International Journal of Modern Physics B}
  \textbf{\bibinfo{volume}{3}}, \bibinfo{pages}{1943} (\bibinfo{year}{1989}).

\bibitem[{\citenamefont{Canright et~al.}(1989)\citenamefont{Canright, Girvin,
  and Brass}}]{Canright89a}
\bibinfo{author}{\bibfnamefont{G.~S.} \bibnamefont{Canright}},
  \bibinfo{author}{\bibfnamefont{S.~M.} \bibnamefont{Girvin}},
  \bibnamefont{and} \bibinfo{author}{\bibfnamefont{A.}~\bibnamefont{Brass}},
  \bibinfo{journal}{Phys. Rev. Lett.} \textbf{\bibinfo{volume}{63}},
  \bibinfo{pages}{2295} (\bibinfo{year}{1989}),
  \urlprefix\url{https://link.aps.org/doi/10.1103/PhysRevLett.63.2295}.

\bibitem[{\citenamefont{Hanna et~al.}(1989)\citenamefont{Hanna, Laughlin, and
  Fetter}}]{Hanna89}
\bibinfo{author}{\bibfnamefont{C.~B.} \bibnamefont{Hanna}},
  \bibinfo{author}{\bibfnamefont{R.~B.} \bibnamefont{Laughlin}},
  \bibnamefont{and} \bibinfo{author}{\bibfnamefont{A.~L.}
  \bibnamefont{Fetter}}, \bibinfo{journal}{Phys. Rev. B}
  \textbf{\bibinfo{volume}{40}}, \bibinfo{pages}{8745} (\bibinfo{year}{1989}),
  \urlprefix\url{https://link.aps.org/doi/10.1103/PhysRevB.40.8745}.

\bibitem[{\citenamefont{Xie et~al.}(1990)\citenamefont{Xie, He, and
  Das~Sarma}}]{Xie90}
\bibinfo{author}{\bibfnamefont{X.~C.} \bibnamefont{Xie}},
  \bibinfo{author}{\bibfnamefont{S.}~\bibnamefont{He}}, \bibnamefont{and}
  \bibinfo{author}{\bibfnamefont{S.}~\bibnamefont{Das~Sarma}},
  \bibinfo{journal}{Phys. Rev. Lett.} \textbf{\bibinfo{volume}{65}},
  \bibinfo{pages}{649} (\bibinfo{year}{1990}),
  \urlprefix\url{https://link.aps.org/doi/10.1103/PhysRevLett.65.649}.

\bibitem[{\citenamefont{Hatsugai et~al.}(1991)\citenamefont{Hatsugai, Kohmoto,
  and Wu}}]{Hatsugai91}
\bibinfo{author}{\bibfnamefont{Y.}~\bibnamefont{Hatsugai}},
  \bibinfo{author}{\bibfnamefont{M.}~\bibnamefont{Kohmoto}}, \bibnamefont{and}
  \bibinfo{author}{\bibfnamefont{Y.-S.} \bibnamefont{Wu}},
  \bibinfo{journal}{Phys. Rev. B} \textbf{\bibinfo{volume}{43}},
  \bibinfo{pages}{10761} (\bibinfo{year}{1991}),
  \urlprefix\url{https://link.aps.org/doi/10.1103/PhysRevB.43.10761}.

\bibitem[{\citenamefont{Kudo and Hatsugai}(2020)}]{Kudo20}
\bibinfo{author}{\bibfnamefont{K.}~\bibnamefont{Kudo}} \bibnamefont{and}
  \bibinfo{author}{\bibfnamefont{Y.}~\bibnamefont{Hatsugai}},
  \bibinfo{journal}{Phys. Rev. B} \textbf{\bibinfo{volume}{102}},
  \bibinfo{pages}{125108} (\bibinfo{year}{2020}),
  \urlprefix\url{https://link.aps.org/doi/10.1103/PhysRevB.102.125108}.

\bibitem[{\citenamefont{Girvin et~al.}(1990)\citenamefont{Girvin, MacDonald,
  Fisher, Rey, and Sethna}}]{Girvin90}
\bibinfo{author}{\bibfnamefont{S.~M.} \bibnamefont{Girvin}},
  \bibinfo{author}{\bibfnamefont{A.~H.} \bibnamefont{MacDonald}},
  \bibinfo{author}{\bibfnamefont{M.~P.~A.} \bibnamefont{Fisher}},
  \bibinfo{author}{\bibfnamefont{S.-J.} \bibnamefont{Rey}}, \bibnamefont{and}
  \bibinfo{author}{\bibfnamefont{J.~P.} \bibnamefont{Sethna}},
  \bibinfo{journal}{Phys. Rev. Lett.} \textbf{\bibinfo{volume}{65}},
  \bibinfo{pages}{1671} (\bibinfo{year}{1990}),
  \urlprefix\url{https://link.aps.org/doi/10.1103/PhysRevLett.65.1671}.

\bibitem[{\citenamefont{Chin and Hu}(1992)}]{Chin92}
\bibinfo{author}{\bibfnamefont{S.~A.} \bibnamefont{Chin}} \bibnamefont{and}
  \bibinfo{author}{\bibfnamefont{C.-R.} \bibnamefont{Hu}},
  \bibinfo{journal}{Phys. Rev. Lett.} \textbf{\bibinfo{volume}{69}},
  \bibinfo{pages}{229} (\bibinfo{year}{1992}),
  \urlprefix\url{https://link.aps.org/doi/10.1103/PhysRevLett.69.229}.

\bibitem[{\citenamefont{Lundholm}(2017)}]{Lundholm17}
\bibinfo{author}{\bibfnamefont{D.}~\bibnamefont{Lundholm}},
  \bibinfo{journal}{Physical Review A} \textbf{\bibinfo{volume}{96}},
  \bibinfo{pages}{012116} (\bibinfo{year}{2017}).

\bibitem[{\citenamefont{Lundholm and Solovej}(2013{\natexlab{a}})}]{Lundholm13}
\bibinfo{author}{\bibfnamefont{D.}~\bibnamefont{Lundholm}} \bibnamefont{and}
  \bibinfo{author}{\bibfnamefont{J.~P.} \bibnamefont{Solovej}},
  \bibinfo{journal}{Physical Review A} \textbf{\bibinfo{volume}{88}},
  \bibinfo{pages}{062106} (\bibinfo{year}{2013}{\natexlab{a}}).

\bibitem[{\citenamefont{Ouvry}(2007)}]{Ouvry07}
\bibinfo{author}{\bibfnamefont{S.}~\bibnamefont{Ouvry}},
  \bibinfo{journal}{arXiv preprint arXiv:0712.2174}  (\bibinfo{year}{2007}).

\bibitem[{\citenamefont{Lundholm and
  Solovej}(2013{\natexlab{b}})}]{Lundholm13a}
\bibinfo{author}{\bibfnamefont{D.}~\bibnamefont{Lundholm}} \bibnamefont{and}
  \bibinfo{author}{\bibfnamefont{J.~P.} \bibnamefont{Solovej}},
  \bibinfo{journal}{Communications in Mathematical Physics}
  \textbf{\bibinfo{volume}{322}}, \bibinfo{pages}{883}
  (\bibinfo{year}{2013}{\natexlab{b}}).

\bibitem[{\citenamefont{Ilieva and Thirring}(2001)}]{Ilieva01}
\bibinfo{author}{\bibfnamefont{N.}~\bibnamefont{Ilieva}} \bibnamefont{and}
  \bibinfo{author}{\bibfnamefont{W.}~\bibnamefont{Thirring}},
  \bibinfo{journal}{Physics Letters B} \textbf{\bibinfo{volume}{504}},
  \bibinfo{pages}{201} (\bibinfo{year}{2001}).

\bibitem[{\citenamefont{Laughlin}(1988)}]{Laughlin88science}
\bibinfo{author}{\bibfnamefont{R.}~\bibnamefont{Laughlin}},
  \bibinfo{journal}{Science} \textbf{\bibinfo{volume}{242}},
  \bibinfo{pages}{525} (\bibinfo{year}{1988}).

\bibitem[{\citenamefont{Fetter et~al.}(1989)\citenamefont{Fetter, Hanna, and
  Laughlin}}]{Fetter89}
\bibinfo{author}{\bibfnamefont{A.}~\bibnamefont{Fetter}},
  \bibinfo{author}{\bibfnamefont{C.}~\bibnamefont{Hanna}}, \bibnamefont{and}
  \bibinfo{author}{\bibfnamefont{R.}~\bibnamefont{Laughlin}},
  \bibinfo{journal}{Physical Review B} \textbf{\bibinfo{volume}{39}},
  \bibinfo{pages}{9679} (\bibinfo{year}{1989}).

\bibitem[{\citenamefont{Wen and Zee}(1990)}]{Wen90b}
\bibinfo{author}{\bibfnamefont{X.}~\bibnamefont{Wen}} \bibnamefont{and}
  \bibinfo{author}{\bibfnamefont{A.}~\bibnamefont{Zee}},
  \bibinfo{journal}{Physical Review B} \textbf{\bibinfo{volume}{41}},
  \bibinfo{pages}{240} (\bibinfo{year}{1990}).

\bibitem[{\citenamefont{Lee}(1991)}]{Lee91}
\bibinfo{author}{\bibfnamefont{D.-H.} \bibnamefont{Lee}},
  \bibinfo{journal}{International Journal of Modern Physics B}
  \textbf{\bibinfo{volume}{5}}, \bibinfo{pages}{1695} (\bibinfo{year}{1991}).

\bibitem[{\citenamefont{Chitra and Sen}(1992)}]{Chitra92}
\bibinfo{author}{\bibfnamefont{R.}~\bibnamefont{Chitra}} \bibnamefont{and}
  \bibinfo{author}{\bibfnamefont{D.}~\bibnamefont{Sen}},
  \bibinfo{journal}{Phys. Rev. B} \textbf{\bibinfo{volume}{46}},
  \bibinfo{pages}{10923} (\bibinfo{year}{1992}),
  \urlprefix\url{https://link.aps.org/doi/10.1103/PhysRevB.46.10923}.

\bibitem[{\citenamefont{Li et~al.}(1992)\citenamefont{Li, Bhaduri, and
  Murthy}}]{Li92}
\bibinfo{author}{\bibfnamefont{S.}~\bibnamefont{Li}},
  \bibinfo{author}{\bibfnamefont{R.~K.} \bibnamefont{Bhaduri}},
  \bibnamefont{and} \bibinfo{author}{\bibfnamefont{M.~V.~N.}
  \bibnamefont{Murthy}}, \bibinfo{journal}{Phys. Rev. B}
  \textbf{\bibinfo{volume}{46}}, \bibinfo{pages}{1228} (\bibinfo{year}{1992}),
  \urlprefix\url{https://link.aps.org/doi/10.1103/PhysRevB.46.1228}.

\bibitem[{\citenamefont{Correggi et~al.}(2017)\citenamefont{Correggi, Lundholm,
  and Rougerie}}]{Correggi17}
\bibinfo{author}{\bibfnamefont{M.}~\bibnamefont{Correggi}},
  \bibinfo{author}{\bibfnamefont{D.}~\bibnamefont{Lundholm}}, \bibnamefont{and}
  \bibinfo{author}{\bibfnamefont{N.}~\bibnamefont{Rougerie}},
  \bibinfo{journal}{Analysis \& PDE} \textbf{\bibinfo{volume}{10}},
  \bibinfo{pages}{1169} (\bibinfo{year}{2017}).

\bibitem[{\citenamefont{Tsui et~al.}(1982)\citenamefont{Tsui, Stormer, and
  Gossard}}]{Tsui82}
\bibinfo{author}{\bibfnamefont{D.~C.} \bibnamefont{Tsui}},
  \bibinfo{author}{\bibfnamefont{H.~L.} \bibnamefont{Stormer}},
  \bibnamefont{and} \bibinfo{author}{\bibfnamefont{A.~C.}
  \bibnamefont{Gossard}}, \bibinfo{journal}{Phys. Rev. Lett.}
  \textbf{\bibinfo{volume}{48}}, \bibinfo{pages}{1559} (\bibinfo{year}{1982}),
  \urlprefix\url{http://link.aps.org/doi/10.1103/PhysRevLett.48.1559}.

\bibitem[{\citenamefont{Nakamura et~al.}(2020)\citenamefont{Nakamura, Liang,
  Gardner, and Manfra}}]{Nakamura20}
\bibinfo{author}{\bibfnamefont{J.}~\bibnamefont{Nakamura}},
  \bibinfo{author}{\bibfnamefont{S.}~\bibnamefont{Liang}},
  \bibinfo{author}{\bibfnamefont{G.~C.} \bibnamefont{Gardner}},
  \bibnamefont{and} \bibinfo{author}{\bibfnamefont{M.~J.}
  \bibnamefont{Manfra}}, \emph{\bibinfo{title}{Direct observation of anyonic
  braiding statistics at the $\nu=1/3$ fractional quantum {Hall} state}}
  (\bibinfo{year}{2020}), \eprint{2006.14115}.

\bibitem[{\citenamefont{Bartolomei et~al.}(2020)\citenamefont{Bartolomei,
  Kumar, Bisognin, Marguerite, Berroir, Bocquillon, Pla{\c c}ais, Cavanna,
  Dong, Gennser et~al.}}]{Bartolomei20}
\bibinfo{author}{\bibfnamefont{H.}~\bibnamefont{Bartolomei}},
  \bibinfo{author}{\bibfnamefont{M.}~\bibnamefont{Kumar}},
  \bibinfo{author}{\bibfnamefont{R.}~\bibnamefont{Bisognin}},
  \bibinfo{author}{\bibfnamefont{A.}~\bibnamefont{Marguerite}},
  \bibinfo{author}{\bibfnamefont{J.-M.} \bibnamefont{Berroir}},
  \bibinfo{author}{\bibfnamefont{E.}~\bibnamefont{Bocquillon}},
  \bibinfo{author}{\bibfnamefont{B.}~\bibnamefont{Pla{\c c}ais}},
  \bibinfo{author}{\bibfnamefont{A.}~\bibnamefont{Cavanna}},
  \bibinfo{author}{\bibfnamefont{Q.}~\bibnamefont{Dong}},
  \bibinfo{author}{\bibfnamefont{U.}~\bibnamefont{Gennser}},
  \bibnamefont{et~al.}, \bibinfo{journal}{Science}
  \textbf{\bibinfo{volume}{368}}, \bibinfo{pages}{173} (\bibinfo{year}{2020}),
  ISSN \bibinfo{issn}{0036-8075},
  \eprint{https://science.sciencemag.org/content/368/6487/173.full.pdf},
  \urlprefix\url{https://science.sciencemag.org/content/368/6487/173}.

\bibitem[{\citenamefont{Halperin}(1984)}]{Halperin84}
\bibinfo{author}{\bibfnamefont{B.~I.} \bibnamefont{Halperin}},
  \bibinfo{journal}{Phys. Rev. Lett.} \textbf{\bibinfo{volume}{52}},
  \bibinfo{pages}{1583} (\bibinfo{year}{1984}),
  \urlprefix\url{http://link.aps.org/doi/10.1103/PhysRevLett.52.1583}.

\bibitem[{\citenamefont{Arovas et~al.}(1985)\citenamefont{Arovas, Schrieffer,
  Wilczek, and Zee}}]{Arovas85}
\bibinfo{author}{\bibfnamefont{D.~P.} \bibnamefont{Arovas}},
  \bibinfo{author}{\bibfnamefont{R.}~\bibnamefont{Schrieffer}},
  \bibinfo{author}{\bibfnamefont{F.}~\bibnamefont{Wilczek}}, \bibnamefont{and}
  \bibinfo{author}{\bibfnamefont{A.}~\bibnamefont{Zee}},
  \bibinfo{journal}{Nuclear Physics B} \textbf{\bibinfo{volume}{251}},
  \bibinfo{pages}{117} (\bibinfo{year}{1985}).

\bibitem[{\citenamefont{Jain}(1989)}]{Jain89}
\bibinfo{author}{\bibfnamefont{J.~K.} \bibnamefont{Jain}},
  \bibinfo{journal}{Phys. Rev. Lett.} \textbf{\bibinfo{volume}{63}},
  \bibinfo{pages}{199} (\bibinfo{year}{1989}),
  \urlprefix\url{http://link.aps.org/doi/10.1103/PhysRevLett.63.199}.

\bibitem[{\citenamefont{Jain}(2007)}]{Jain07}
\bibinfo{author}{\bibfnamefont{J.~K.} \bibnamefont{Jain}},
  \emph{\bibinfo{title}{Composite Fermions}} (\bibinfo{publisher}{Cambridge
  University Press, New York, US}, \bibinfo{year}{2007}).

\bibitem[{\citenamefont{Halperin and Jain}(2020)}]{Halperin20}
\bibinfo{editor}{\bibfnamefont{B.~I.} \bibnamefont{Halperin}} \bibnamefont{and}
  \bibinfo{editor}{\bibfnamefont{J.~K.} \bibnamefont{Jain}}, eds.,
  \emph{\bibinfo{title}{{Fractional} {Quantum} {Hall} {Effects} {New}
  {Developments}}} (\bibinfo{publisher}{World Scientific},
  \bibinfo{year}{2020}),
  \eprint{https://worldscientific.com/doi/pdf/10.1142/11751},
  \urlprefix\url{https://worldscientific.com/doi/abs/10.1142/11751}.

\bibitem[{\citenamefont{Kj{\o}nsberg and Leinaas}(1999)}]{Kjonsberg99b}
\bibinfo{author}{\bibfnamefont{H.}~\bibnamefont{Kj{\o}nsberg}}
  \bibnamefont{and} \bibinfo{author}{\bibfnamefont{J.~M.}
  \bibnamefont{Leinaas}}, \bibinfo{journal}{Nucl. Phys. B}
  \textbf{\bibinfo{volume}{559}}, \bibinfo{pages}{705} (\bibinfo{year}{1999}).

\bibitem[{\citenamefont{Jeon et~al.}(2003)\citenamefont{Jeon, Graham, and
  Jain}}]{Jeon03b}
\bibinfo{author}{\bibfnamefont{G.~S.} \bibnamefont{Jeon}},
  \bibinfo{author}{\bibfnamefont{K.~L.} \bibnamefont{Graham}},
  \bibnamefont{and} \bibinfo{author}{\bibfnamefont{J.~K.} \bibnamefont{Jain}},
  \bibinfo{journal}{Phys. Rev. Lett.} \textbf{\bibinfo{volume}{91}},
  \bibinfo{pages}{036801} (\bibinfo{year}{2003}),
  \urlprefix\url{http://link.aps.org/doi/10.1103/PhysRevLett.91.036801}.

\bibitem[{\citenamefont{Jeon et~al.}(2004)\citenamefont{Jeon, Graham, and
  Jain}}]{Jeon04}
\bibinfo{author}{\bibfnamefont{G.~S.} \bibnamefont{Jeon}},
  \bibinfo{author}{\bibfnamefont{K.~L.} \bibnamefont{Graham}},
  \bibnamefont{and} \bibinfo{author}{\bibfnamefont{J.~K.} \bibnamefont{Jain}},
  \bibinfo{journal}{Phys. Rev. B} \textbf{\bibinfo{volume}{70}},
  \bibinfo{pages}{125316} (\bibinfo{year}{2004}),
  \urlprefix\url{http://link.aps.org/doi/10.1103/PhysRevB.70.125316}.

\bibitem[{\citenamefont{Hu and Jain}(2019)}]{Hu19}
\bibinfo{author}{\bibfnamefont{Y.}~\bibnamefont{Hu}} \bibnamefont{and}
  \bibinfo{author}{\bibfnamefont{J.~K.} \bibnamefont{Jain}},
  \bibinfo{journal}{Phys. Rev. Lett.} \textbf{\bibinfo{volume}{123}},
  \bibinfo{pages}{176802} (\bibinfo{year}{2019}),
  \urlprefix\url{https://link.aps.org/doi/10.1103/PhysRevLett.123.176802}.

\bibitem[{\citenamefont{Lee and Fisher}(1989)}]{Lee89}
\bibinfo{author}{\bibfnamefont{D.-H.} \bibnamefont{Lee}} \bibnamefont{and}
  \bibinfo{author}{\bibfnamefont{M.~P.} \bibnamefont{Fisher}},
  \bibinfo{journal}{Physical review letters} \textbf{\bibinfo{volume}{63}},
  \bibinfo{pages}{903} (\bibinfo{year}{1989}).

\bibitem[{\citenamefont{McMullen et~al.}(1991)\citenamefont{McMullen, Jena, and
  Khanna}}]{Mcmullen91}
\bibinfo{author}{\bibfnamefont{T.}~\bibnamefont{McMullen}},
  \bibinfo{author}{\bibfnamefont{P.}~\bibnamefont{Jena}}, \bibnamefont{and}
  \bibinfo{author}{\bibfnamefont{S.}~\bibnamefont{Khanna}},
  \bibinfo{journal}{International Journal of Modern Physics B}
  \textbf{\bibinfo{volume}{5}}, \bibinfo{pages}{1579} (\bibinfo{year}{1991}).

\bibitem[{\citenamefont{de~Gennes}(1999)}]{Pierre99}
\bibinfo{author}{\bibfnamefont{P.-G.} \bibnamefont{de~Gennes}},
  \emph{\bibinfo{title}{Superconductivity of Metals and Alloys}}, Advanced book
  classics (\bibinfo{publisher}{Advanced Book Program, Perseus Books},
  \bibinfo{year}{1999}), ISBN \bibinfo{isbn}{9780738201016},
  \urlprefix\url{https://books.google.com/books?id=xacsAAAAYAAJ}.

\bibitem[{\citenamefont{Ma and Zhang}(1991)}]{Ma91}
\bibinfo{author}{\bibfnamefont{M.}~\bibnamefont{Ma}} \bibnamefont{and}
  \bibinfo{author}{\bibfnamefont{F.~C.} \bibnamefont{Zhang}},
  \bibinfo{journal}{Phys. Rev. Lett.} \textbf{\bibinfo{volume}{66}},
  \bibinfo{pages}{1769} (\bibinfo{year}{1991}),
  \urlprefix\url{https://link.aps.org/doi/10.1103/PhysRevLett.66.1769}.

\bibitem[{\citenamefont{Hohenberg and Kohn}(1964)}]{Hohenberg64}
\bibinfo{author}{\bibfnamefont{P.}~\bibnamefont{Hohenberg}} \bibnamefont{and}
  \bibinfo{author}{\bibfnamefont{W.}~\bibnamefont{Kohn}},
  \bibinfo{journal}{Phys. Rev.} \textbf{\bibinfo{volume}{136}},
  \bibinfo{pages}{B864} (\bibinfo{year}{1964}),
  \urlprefix\url{https://link.aps.org/doi/10.1103/PhysRev.136.B864}.

\bibitem[{\citenamefont{Kohn and Sham}(1965)}]{Kohn65}
\bibinfo{author}{\bibfnamefont{W.}~\bibnamefont{Kohn}} \bibnamefont{and}
  \bibinfo{author}{\bibfnamefont{L.~J.} \bibnamefont{Sham}},
  \bibinfo{journal}{Phys. Rev.} \textbf{\bibinfo{volume}{140}},
  \bibinfo{pages}{A1133} (\bibinfo{year}{1965}),
  \urlprefix\url{https://link.aps.org/doi/10.1103/PhysRev.140.A1133}.

\bibitem[{\citenamefont{Giuliani and Vignale}(2008)}]{Giuliani08}
\bibinfo{author}{\bibfnamefont{G.}~\bibnamefont{Giuliani}} \bibnamefont{and}
  \bibinfo{author}{\bibfnamefont{G.}~\bibnamefont{Vignale}},
  \emph{\bibinfo{title}{Quantum Theory of the Electron Liquid}}
  (\bibinfo{publisher}{Cambridge University Press, The Edinburgh Building,
  Cambridge CB2 2RU, UK}, \bibinfo{year}{2008}).

\bibitem[{\citenamefont{Ferconi et~al.}(1995)\citenamefont{Ferconi, Geller, and
  Vignale}}]{Ferconi95}
\bibinfo{author}{\bibfnamefont{M.}~\bibnamefont{Ferconi}},
  \bibinfo{author}{\bibfnamefont{M.~R.} \bibnamefont{Geller}},
  \bibnamefont{and} \bibinfo{author}{\bibfnamefont{G.}~\bibnamefont{Vignale}},
  \bibinfo{journal}{Phys. Rev. B} \textbf{\bibinfo{volume}{52}},
  \bibinfo{pages}{16357} (\bibinfo{year}{1995}),
  \urlprefix\url{http://link.aps.org/doi/10.1103/PhysRevB.52.16357}.

\bibitem[{\citenamefont{Heinonen et~al.}(1995)\citenamefont{Heinonen, Lubin,
  and Johnson}}]{Heinonen95}
\bibinfo{author}{\bibfnamefont{O.}~\bibnamefont{Heinonen}},
  \bibinfo{author}{\bibfnamefont{M.~I.} \bibnamefont{Lubin}}, \bibnamefont{and}
  \bibinfo{author}{\bibfnamefont{M.~D.} \bibnamefont{Johnson}},
  \bibinfo{journal}{Phys. Rev. Lett.} \textbf{\bibinfo{volume}{75}},
  \bibinfo{pages}{4110} (\bibinfo{year}{1995}),
  \urlprefix\url{http://link.aps.org/doi/10.1103/PhysRevLett.75.4110}.

\bibitem[{\citenamefont{Zhao et~al.}(2017)\citenamefont{Zhao, Thakurathi, Jain,
  Sen, and Jain}}]{Zhao17}
\bibinfo{author}{\bibfnamefont{J.}~\bibnamefont{Zhao}},
  \bibinfo{author}{\bibfnamefont{M.}~\bibnamefont{Thakurathi}},
  \bibinfo{author}{\bibfnamefont{M.}~\bibnamefont{Jain}},
  \bibinfo{author}{\bibfnamefont{D.}~\bibnamefont{Sen}}, \bibnamefont{and}
  \bibinfo{author}{\bibfnamefont{J.~K.} \bibnamefont{Jain}},
  \bibinfo{journal}{Phys. Rev. Lett.} \textbf{\bibinfo{volume}{118}},
  \bibinfo{pages}{196802} (\bibinfo{year}{2017}),
  \urlprefix\url{https://link.aps.org/doi/10.1103/PhysRevLett.118.196802}.

\bibitem[{\citenamefont{Laughlin}(1983)}]{Laughlin83}
\bibinfo{author}{\bibfnamefont{R.~B.} \bibnamefont{Laughlin}},
  \bibinfo{journal}{Phys. Rev. Lett.} \textbf{\bibinfo{volume}{50}},
  \bibinfo{pages}{1395} (\bibinfo{year}{1983}),
  \urlprefix\url{http://link.aps.org/doi/10.1103/PhysRevLett.50.1395}.

\bibitem[{\citenamefont{Sen and Chitra}(1992)}]{Diptiman92b}
\bibinfo{author}{\bibfnamefont{D.}~\bibnamefont{Sen}} \bibnamefont{and}
  \bibinfo{author}{\bibfnamefont{R.}~\bibnamefont{Chitra}},
  \bibinfo{journal}{Phys. Rev. B} \textbf{\bibinfo{volume}{45}},
  \bibinfo{pages}{881} (\bibinfo{year}{1992}),
  \urlprefix\url{https://link.aps.org/doi/10.1103/PhysRevB.45.881}.

\bibitem[{\citenamefont{Mancarella et~al.}(2013)\citenamefont{Mancarella,
  Trombettoni, and Mussardo}}]{Mancarella13}
\bibinfo{author}{\bibfnamefont{F.}~\bibnamefont{Mancarella}},
  \bibinfo{author}{\bibfnamefont{A.}~\bibnamefont{Trombettoni}},
  \bibnamefont{and} \bibinfo{author}{\bibfnamefont{G.}~\bibnamefont{Mussardo}},
  \bibinfo{journal}{Nuclear Physics B} \textbf{\bibinfo{volume}{867}},
  \bibinfo{pages}{950} (\bibinfo{year}{2013}).

\bibitem[{\citenamefont{Grayce and Harris}(1994)}]{Grayce94}
\bibinfo{author}{\bibfnamefont{C.~J.} \bibnamefont{Grayce}} \bibnamefont{and}
  \bibinfo{author}{\bibfnamefont{R.~A.} \bibnamefont{Harris}},
  \bibinfo{journal}{Physical Review A} \textbf{\bibinfo{volume}{50}},
  \bibinfo{pages}{3089} (\bibinfo{year}{1994}).

\bibitem[{\citenamefont{Kohn et~al.}(2004)\citenamefont{Kohn, Savin, and
  Ullrich}}]{kohn04}
\bibinfo{author}{\bibfnamefont{W.}~\bibnamefont{Kohn}},
  \bibinfo{author}{\bibfnamefont{A.}~\bibnamefont{Savin}}, \bibnamefont{and}
  \bibinfo{author}{\bibfnamefont{C.~A.} \bibnamefont{Ullrich}},
  \bibinfo{journal}{International journal of quantum chemistry}
  \textbf{\bibinfo{volume}{100}}, \bibinfo{pages}{20} (\bibinfo{year}{2004}).

\bibitem[{\citenamefont{Tellgren et~al.}(2012)\citenamefont{Tellgren, Kvaal,
  Sagvolden, Ekstr\"om, Teale, and Helgaker}}]{Tellgren12}
\bibinfo{author}{\bibfnamefont{E.~I.} \bibnamefont{Tellgren}},
  \bibinfo{author}{\bibfnamefont{S.}~\bibnamefont{Kvaal}},
  \bibinfo{author}{\bibfnamefont{E.}~\bibnamefont{Sagvolden}},
  \bibinfo{author}{\bibfnamefont{U.}~\bibnamefont{Ekstr\"om}},
  \bibinfo{author}{\bibfnamefont{A.~M.} \bibnamefont{Teale}}, \bibnamefont{and}
  \bibinfo{author}{\bibfnamefont{T.}~\bibnamefont{Helgaker}},
  \bibinfo{journal}{Phys. Rev. A} \textbf{\bibinfo{volume}{86}},
  \bibinfo{pages}{062506} (\bibinfo{year}{2012}),
  \urlprefix\url{https://link.aps.org/doi/10.1103/PhysRevA.86.062506}.

\bibitem[{\citenamefont{Tellgren et~al.}(2018)\citenamefont{Tellgren,
  Laestadius, Helgaker, Kvaal, and Teale}}]{Tellgren18b}
\bibinfo{author}{\bibfnamefont{E.~I.} \bibnamefont{Tellgren}},
  \bibinfo{author}{\bibfnamefont{A.}~\bibnamefont{Laestadius}},
  \bibinfo{author}{\bibfnamefont{T.}~\bibnamefont{Helgaker}},
  \bibinfo{author}{\bibfnamefont{S.}~\bibnamefont{Kvaal}}, \bibnamefont{and}
  \bibinfo{author}{\bibfnamefont{A.~M.} \bibnamefont{Teale}},
  \bibinfo{journal}{The Journal of chemical physics}
  \textbf{\bibinfo{volume}{148}}, \bibinfo{pages}{024101}
  (\bibinfo{year}{2018}).

\bibitem[{\citenamefont{Levy}(1979)}]{Levy79}
\bibinfo{author}{\bibfnamefont{M.}~\bibnamefont{Levy}},
  \bibinfo{journal}{Proceedings of the National Academy of Sciences}
  \textbf{\bibinfo{volume}{76}}, \bibinfo{pages}{6062} (\bibinfo{year}{1979}).

\bibitem[{\citenamefont{Lieb}(1983)}]{Lieb83}
\bibinfo{author}{\bibfnamefont{E.~H.} \bibnamefont{Lieb}},
  \bibinfo{journal}{Int. J. Quantum Chem.} \textbf{\bibinfo{volume}{24}},
  \bibinfo{pages}{243} (\bibinfo{year}{1983}).

\bibitem[{\citenamefont{Vignale and Rasolt}(1987)}]{Vignale87}
\bibinfo{author}{\bibfnamefont{G.}~\bibnamefont{Vignale}} \bibnamefont{and}
  \bibinfo{author}{\bibfnamefont{M.}~\bibnamefont{Rasolt}},
  \bibinfo{journal}{Phys. Rev. Lett.} \textbf{\bibinfo{volume}{59}},
  \bibinfo{pages}{2360} (\bibinfo{year}{1987}),
  \urlprefix\url{https://link.aps.org/doi/10.1103/PhysRevLett.59.2360}.

\bibitem[{\citenamefont{Pribram-Jones et~al.}(2014)\citenamefont{Pribram-Jones,
  Pittalis, Gross, and Burke}}]{Jones14}
\bibinfo{author}{\bibfnamefont{A.}~\bibnamefont{Pribram-Jones}},
  \bibinfo{author}{\bibfnamefont{S.}~\bibnamefont{Pittalis}},
  \bibinfo{author}{\bibfnamefont{E.}~\bibnamefont{Gross}}, \bibnamefont{and}
  \bibinfo{author}{\bibfnamefont{K.}~\bibnamefont{Burke}}, in
  \emph{\bibinfo{booktitle}{Frontiers and Challenges in Warm Dense Matter}}
  (\bibinfo{publisher}{Springer}, \bibinfo{year}{2014}), pp.
  \bibinfo{pages}{25--60}.

\bibitem[{\citenamefont{Choi et~al.}(1999)\citenamefont{Choi, Lee, and
  Ryu}}]{Choi99}
\bibinfo{author}{\bibfnamefont{T.}~\bibnamefont{Choi}},
  \bibinfo{author}{\bibfnamefont{J.}~\bibnamefont{Lee}}, \bibnamefont{and}
  \bibinfo{author}{\bibfnamefont{C.-M.} \bibnamefont{Ryu}},
  \bibinfo{journal}{Modern Physics Letters B} \textbf{\bibinfo{volume}{13}},
  \bibinfo{pages}{925} (\bibinfo{year}{1999}).

\bibitem[{\citenamefont{Wu}(1984)}]{Wu84}
\bibinfo{author}{\bibfnamefont{Y.-S.} \bibnamefont{Wu}},
  \bibinfo{journal}{Phys. Rev. Lett.} \textbf{\bibinfo{volume}{53}},
  \bibinfo{pages}{111} (\bibinfo{year}{1984}),
  \urlprefix\url{https://link.aps.org/doi/10.1103/PhysRevLett.53.111}.

\bibitem[{\citenamefont{Myrheim}(1999)}]{Myrheim99}
\bibinfo{author}{\bibfnamefont{J.}~\bibnamefont{Myrheim}}, in
  \emph{\bibinfo{booktitle}{Aspects topologiques de la physique en basse
  dimension. Topological aspects of low dimensional systems}}
  (\bibinfo{publisher}{Springer}, \bibinfo{year}{1999}), pp.
  \bibinfo{pages}{265--413}.

\bibitem[{\citenamefont{Kaduk et~al.}(2011)\citenamefont{Kaduk, Kowalczyk, and
  Van~Voorhis}}]{Kaduk12}
\bibinfo{author}{\bibfnamefont{B.}~\bibnamefont{Kaduk}},
  \bibinfo{author}{\bibfnamefont{T.}~\bibnamefont{Kowalczyk}},
  \bibnamefont{and}
  \bibinfo{author}{\bibfnamefont{T.}~\bibnamefont{Van~Voorhis}},
  \bibinfo{journal}{Chemical reviews} \textbf{\bibinfo{volume}{112}},
  \bibinfo{pages}{321} (\bibinfo{year}{2011}).

\bibitem[{\citenamefont{Butts and Rokhsar}(1997)}]{Butts97}
\bibinfo{author}{\bibfnamefont{D.~A.} \bibnamefont{Butts}} \bibnamefont{and}
  \bibinfo{author}{\bibfnamefont{D.~S.} \bibnamefont{Rokhsar}},
  \bibinfo{journal}{Phys. Rev. A} \textbf{\bibinfo{volume}{55}},
  \bibinfo{pages}{4346} (\bibinfo{year}{1997}),
  \urlprefix\url{https://link.aps.org/doi/10.1103/PhysRevA.55.4346}.

\bibitem[{\citenamefont{Lieb et~al.}(1995)\citenamefont{Lieb, Solovej, and
  Yngvason}}]{Lieb95}
\bibinfo{author}{\bibfnamefont{E.~H.} \bibnamefont{Lieb}},
  \bibinfo{author}{\bibfnamefont{J.~P.} \bibnamefont{Solovej}},
  \bibnamefont{and} \bibinfo{author}{\bibfnamefont{J.}~\bibnamefont{Yngvason}},
  \bibinfo{journal}{Phys. Rev. B} \textbf{\bibinfo{volume}{51}},
  \bibinfo{pages}{10646} (\bibinfo{year}{1995}),
  \urlprefix\url{https://link.aps.org/doi/10.1103/PhysRevB.51.10646}.


\bibitem[{\citenamefont{Laughlin}(1981)}]{Laughlin81}
\bibinfo{author}{\bibfnamefont{R.~B.} \bibnamefont{Laughlin}},
  \bibinfo{journal}{Phys. Rev. B} \textbf{\bibinfo{volume}{23}},
  \bibinfo{pages}{5632} (\bibinfo{year}{1981}),
  \urlprefix\url{http://link.aps.org/doi/10.1103/PhysRevB.23.5632}.

\bibitem[{\citenamefont{Halperin}(1982)}]{Halperin82}
\bibinfo{author}{\bibfnamefont{B.~I.} \bibnamefont{Halperin}},
  \bibinfo{journal}{Phys. Rev. B} \textbf{\bibinfo{volume}{25}},
  \bibinfo{pages}{2185} (\bibinfo{year}{1982}),
  \urlprefix\url{http://link.aps.org/doi/10.1103/PhysRevB.25.2185}.

\bibitem[{\citenamefont{Arovas et~al.}(1984)\citenamefont{Arovas, Schrieffer,
  and Wilczek}}]{Arovas84}
\bibinfo{author}{\bibfnamefont{D.}~\bibnamefont{Arovas}},
  \bibinfo{author}{\bibfnamefont{J.~R.} \bibnamefont{Schrieffer}},
  \bibnamefont{and} \bibinfo{author}{\bibfnamefont{F.}~\bibnamefont{Wilczek}},
  \bibinfo{journal}{Phys. Rev. Lett.} \textbf{\bibinfo{volume}{53}},
  \bibinfo{pages}{722} (\bibinfo{year}{1984}),
  \urlprefix\url{http://link.aps.org/doi/10.1103/PhysRevLett.53.722}.

\bibitem[{\citenamefont{Ciftja et~al.}(2005)\citenamefont{Ciftja, Japaridze,
  and Wang}}]{Ciftja05}
\bibinfo{author}{\bibfnamefont{O.}~\bibnamefont{Ciftja}},
  \bibinfo{author}{\bibfnamefont{G.~S.} \bibnamefont{Japaridze}},
  \bibnamefont{and} \bibinfo{author}{\bibfnamefont{X.-Q.} \bibnamefont{Wang}},
  \bibinfo{journal}{Journal of Physics: Condensed Matter}
  \textbf{\bibinfo{volume}{17}}, \bibinfo{pages}{2977} (\bibinfo{year}{2005}).

\bibitem[{\citenamefont{Byers and Yang}(1961)}]{Byers61}
\bibinfo{author}{\bibfnamefont{N.}~\bibnamefont{Byers}} \bibnamefont{and}
  \bibinfo{author}{\bibfnamefont{C.}~\bibnamefont{Yang}},
  \bibinfo{journal}{Physical review letters} \textbf{\bibinfo{volume}{7}},
  \bibinfo{pages}{46} (\bibinfo{year}{1961}).

\bibitem[{\citenamefont{Chou}(1991)}]{Chou91}
\bibinfo{author}{\bibfnamefont{C.}~\bibnamefont{Chou}}, \bibinfo{journal}{Phys.
  Rev. D} \textbf{\bibinfo{volume}{44}}, \bibinfo{pages}{2533}
  (\bibinfo{year}{1991}),
  \urlprefix\url{https://link.aps.org/doi/10.1103/PhysRevD.44.2533}.

\bibitem[{\citenamefont{Iyetomi and Vashishta}(1989{\natexlab{a}})}]{Iyetomi89}
\bibinfo{author}{\bibfnamefont{H.}~\bibnamefont{Iyetomi}} \bibnamefont{and}
  \bibinfo{author}{\bibfnamefont{P.}~\bibnamefont{Vashishta}},
  \bibinfo{journal}{Solid State Ionics} \textbf{\bibinfo{volume}{32}},
  \bibinfo{pages}{959} (\bibinfo{year}{1989}{\natexlab{a}}).

\bibitem[{\citenamefont{Iyetomi and
  Vashishta}(1989{\natexlab{b}})}]{Iyetomi89a}
\bibinfo{author}{\bibfnamefont{H.}~\bibnamefont{Iyetomi}} \bibnamefont{and}
  \bibinfo{author}{\bibfnamefont{P.}~\bibnamefont{Vashishta}},
  \bibinfo{journal}{Journal of Physics: Condensed Matter}
  \textbf{\bibinfo{volume}{1}}, \bibinfo{pages}{1899}
  (\bibinfo{year}{1989}{\natexlab{b}}).

\bibitem[{\citenamefont{Gezerlis and Bertsch}(2010)}]{Gezerlis10}
\bibinfo{author}{\bibfnamefont{A.}~\bibnamefont{Gezerlis}} \bibnamefont{and}
  \bibinfo{author}{\bibfnamefont{G.}~\bibnamefont{Bertsch}},
  \bibinfo{journal}{Physical review letters} \textbf{\bibinfo{volume}{105}},
  \bibinfo{pages}{212501} (\bibinfo{year}{2010}).

\bibitem[{\citenamefont{Fushiki et~al.}(1992)\citenamefont{Fushiki,
  Gudmundsson, Pethick, and Yngvason}}]{Fushiki92}
\bibinfo{author}{\bibfnamefont{I.}~\bibnamefont{Fushiki}},
  \bibinfo{author}{\bibfnamefont{E.~H.} \bibnamefont{Gudmundsson}},
  \bibinfo{author}{\bibfnamefont{C.~J.} \bibnamefont{Pethick}},
  \bibnamefont{and} \bibinfo{author}{\bibfnamefont{J.}~\bibnamefont{Yngvason}},
  \bibinfo{journal}{Annals of Physics} \textbf{\bibinfo{volume}{216}},
  \bibinfo{pages}{29} (\bibinfo{year}{1992}).

\end{thebibliography}
%\bibliographystyle{apsrev}

\end{document}